\documentclass[prl,twocolumn,amsmath,amssymb,english]{revtex4-1}

\usepackage{natbib}
\usepackage{subfigure}
\usepackage{tabularx}
\usepackage{epsfig}
\usepackage{longtable}
\usepackage{amsfonts}
\usepackage{rotating}
\usepackage{subfigure}
\usepackage{amsmath}
\usepackage{hyperref}
\usepackage{babel}
\usepackage{comment}

\usepackage{color}

\newcommand{\bea}{\begin{eqnarray}}
\newcommand{\eea}{\end{eqnarray}}
\newcommand{\la}{\label}
\newcommand{\be}{\begin{equation}}
\newcommand{\ee}{\end{equation}}

\newcommand{\sgn}{\,\mbox{sgn}}



\begin{document}

\title{Nearest neighbor tight binding models with an exact mobility edge in one dimension}
 \author{Sriram Ganeshan}
 \author{J. H. Pixley}
\author{S. Das Sarma}
\affiliation{Condensed Matter Theory Center and Joint Quantum Institute, Department of Physics, University of Maryland, College Park, MD 20742, USA}

\date{\today}

%%%%%%%%%%%%%
\begin{abstract}
We investigate localization properties in a family of deterministic (i.e. no disorder) nearest neighbor tight binding models with quasiperiodic onsite modulation. We prove that this family is self-dual under a generalized duality transformation. The self-dual condition for this general model turns out to be a simple closed form function of the model parameters and energy. We introduce the typical density of states as an order parameter for localization in quasiperiodic systems. By direct calculations of the inverse participation ratio and the typical density of states we numerically verify that this self-dual line indeed defines a mobility edge in energy separating localized and extended states.  Our model is a first example of a nearest neighbor tight binding model manifesting a mobility edge protected by a duality symmetry. We propose a realistic experimental scheme to realize our results in atomic optical lattices and photonic waveguides.
\end{abstract}
%%%%%%%%%%%%%
\maketitle
%\tableofcontents
Anderson localization~\cite{anderson} is a universal and extensively studied property of a quantum particle or a wave in a disordered medium.  An interesting consequence of Anderson localization is a quantum phase transition between extended and localized states as a function of the disorder strength. In three dimensional systems with random (uncorrelated) disorder, a localization transition occurs as the strength of disorder crosses a critical value forming a sharp energy dependent mobility edge at the phase boundary separating localized/extended states below/above the mobility edge.  Scaling theory~\cite{tvr} has shown the absence of this critical behavior in one and two dimensions where all states, at least in the absence of interaction, are known to be localized in a disordered system, pushing the mobility edge effectively to infinite energy. However, in one dimension this picture changes for a quasiperiodic system with two incommensurate (but deterministic) lattice potentials, which in a loose qualitative sense might be construed to be a highly correlated disorder, albeit perfectly well-defined with no randomness whatsoever. Aubry and Andre~\cite{AA} showed that a 1D tight binding model with an onsite cosine modulation incommensurate with the underlying lattice has a self-dual symmetry and manifests an energy \emph{independent} localization transition as a function of the modulation strength, i.e., all states are either localized or extended depending on the relative strength of the incommensurate modulation potential with respect to the lattice potential. The same model was earlier considered by Harper ~\cite{harper55}  and by Azbel~\cite{azbel} and Hofstadter~\cite{hofstadter} to study the self similar spectrum of conduction electrons in an external magnetic field (we abbreviate this model as AAH from hereon). 

This result~\cite{AA} has led to an extensive theoretical investigation of the AAH model in the context of 
localization~\cite{sankarprl88, thoulessprl88, sankarprb90, biddlepra09, biddleprl10, sankarprl86} in an incommensurate potential. Recent experimental developments in photonic crystals~\cite{lahini2009, kraus1, kraus3, krausfib} and ultracold atoms~\cite{wiersma, roati08,modugno} have led to the implementation of  the quasiperiodic AAH model where this localization transition has been observed. Recent works have used  analytical~\cite{altshuler} and numerical methods~\cite{vadim} to show the existence of a many body localization transition in the quasiperiodic  AAH model in the presence of weak interactions.  The duality-driven and energy independent localization transition in the AAH model does not manifest a mobility edge, which is a hallmark of the disorder tuned localization transition in 3D. The 1D localization transition in the AAH model, defined by the self-duality point, thus has
%has thus
 no analog in 
 %the
  disorder-driven Anderson localization, and does not give any insight into the physics of 3D mobility edges. Recent works have shown the existence of a mobility edge in quasiperiodic flat band models~\cite{flach}.

In this letter we show that there exists a general family of quasiperiodic models with nearest neighbor hopping that  are self-dual under a generalized transformation. We analytically show that this general family has a true 1D mobility edge, not allowed in the Anderson model and the AAH model, which can be expressed as a closed form expression involving energy and system parameters. 
%This self-duality is robust in the sense that as long as  some basic symmetry conditions are satisfied, we can deform this model without destroying self-duality. 
In addition, we introduce the typical density of states (i.e. the \emph{spatial} geometric average of the local density of states) as a generic order parameter for localization in quasi periodic systems that can naturally be generalized to interacting many body problems. 
We organize this letter by first writing the self-duality condition for a specific onsite model. We then provide a physical intuition for this novel self-dual critical condition and present numerical verification for the existence of the mobility edge through calculations of the inverse participation ratio and the typical density of states. We then present a different model satisfying exactly the same critical condition with a totally different energy spectrum. After developing some physical intuition for our result, we explicitly prove that the mobility edge is precisely the self-duality condition for a broad class of nearest neighbor models. We conclude by presenting a concrete schematic to engineer our model in ultracold atoms and photonic waveguides. The experimental observation of a 1D mobility edge would be an exciting and surprising result, leading to a deeper understanding of quantum localization phenomena.
\vspace{0.5cm}

\textit{Model:-} We consider a family of 1D tight binding 
models
 with an onsite modulation $V_n$ defined as
 \begin{equation}
 t (u_{n-1}+u_{n+1})+V_n(\alpha,\phi)u_n=Eu_n.\label{model}
 \end{equation}
 %(Eq.~(\ref{model})). 
 The onsite potential, $V_{n}(\alpha, \phi)$, is characterized by the deformation parameter $\alpha$, onsite modulation strength $\lambda$, period $1/b$ and the phase parameter $\phi$ which is redundant in the context of localization. For a quasiperiodic modulation, we set $b$ to be irrational (we choose $1/b=\frac{\sqrt{5}-1}{2}$ for our numerical work although any other irrational choice for $b$ is equally acceptable). 
 
 The first family of models we consider are specified by an onsite potential
 \begin{equation}
 V_n(\alpha,\phi)=2\lambda\frac{\cos (2\pi n b+\phi)}{1-\alpha\cos (2\pi n b+\phi)}.\la{onsite}
 \end{equation}
%\begin{align}
%t (u_{n-1}+u_{n+1})+V_n(\alpha,\phi)u_n=Eu_n.\label{model}\\
%\text{where}, V_n(\alpha,\phi)=2\lambda\frac{\cos (2\pi n b+\phi)}{1-\alpha\cos (2\pi n b+\phi)}\la{onsite}
%\end{align} 
 This onsite potential is a smooth function of $\alpha$ in the open interval $\alpha \in (-1,1)$. $V_{n}(\alpha,\phi)$ has singularities at $\alpha=\pm1$  which we approach only in a limiting sense.  Each value of $\alpha$ corresponds to a different tight binding model containing the AAH ($\alpha=0$) model as a limiting case and a general quasi periodic model with correlated singularities at $\alpha=\pm1$. 

The AAH limit ($\alpha=0$) with irrational $b$ manifests a localization transition at the self-dual point $\lambda=t$~\cite{AA}. A vast body of numerical work~\cite{sankarprl88, sankarprb90, biddlepra09, biddleprl10, sankarprl86} has been done to understand how this critical point changes once the duality symmetry of the AAH model is broken in some controlled fashion. Based on numerical results, a general consensus has prevailed that the $|\lambda|=|t|$ critical point modifies into a mobility edge. The $\alpha\ne 0$ in Eq.~(\ref{onsite}) can be thought of as a perturbation that breaks this duality symmetry. 
%For $|\alpha| \ll 1$, one can approximate the onsite potential as $V^1_{n}(\alpha,\phi)\sim \cos(2\pi n b+\phi)+\alpha \cos^2(2\pi n b+\phi)$. The first term is the AAH model and the second term can be thought of as a small duality breaking perturbation that results in an $\alpha$ dependent mobility edge. The critical mobility edge for this perturbed model is some unknown  function of $\lambda$, $t$, $E$ and $\alpha$ which can be determined, if necessary, from numerical work. 
%%%%%%%%%%%%%%%%%%%%%%%%%%%%%%%%%%%%%%%%%%%%%%%%%%%%%%%%%%%%%%%%%%%%%%%%%%%%%%%%%%%%%%%%
 \begin{figure}[htb!]
  \centering
\includegraphics[width=4cm,height=3.5cm]{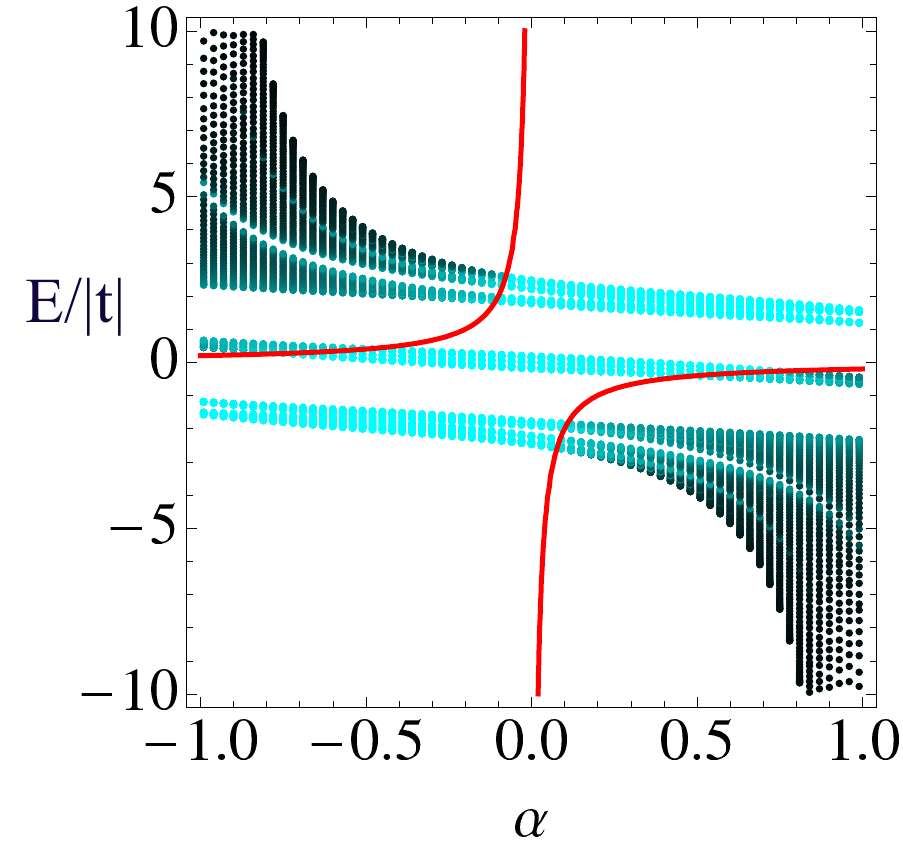}\hspace{0.1cm}\includegraphics[width=4.0cm,height=3.5cm]{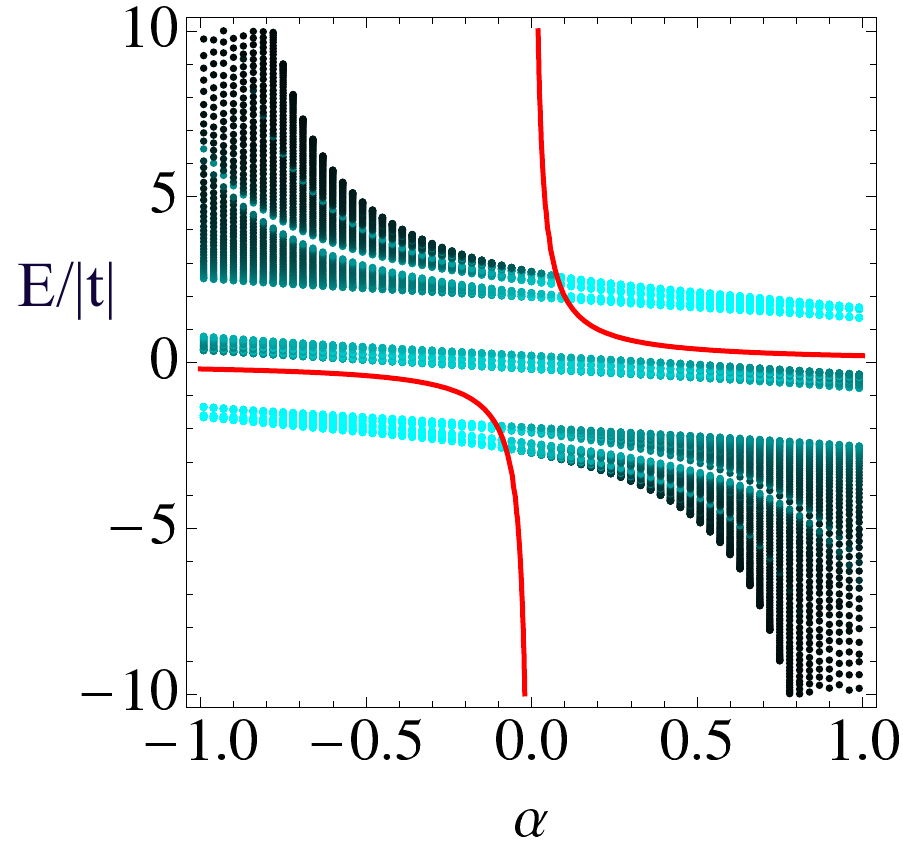}\\
\hspace{0.7cm}\textrm{(a) IPR}\hspace{4cm}\textrm{(b)IPR}\\
\includegraphics[width=4.0cm,height=3.5cm]{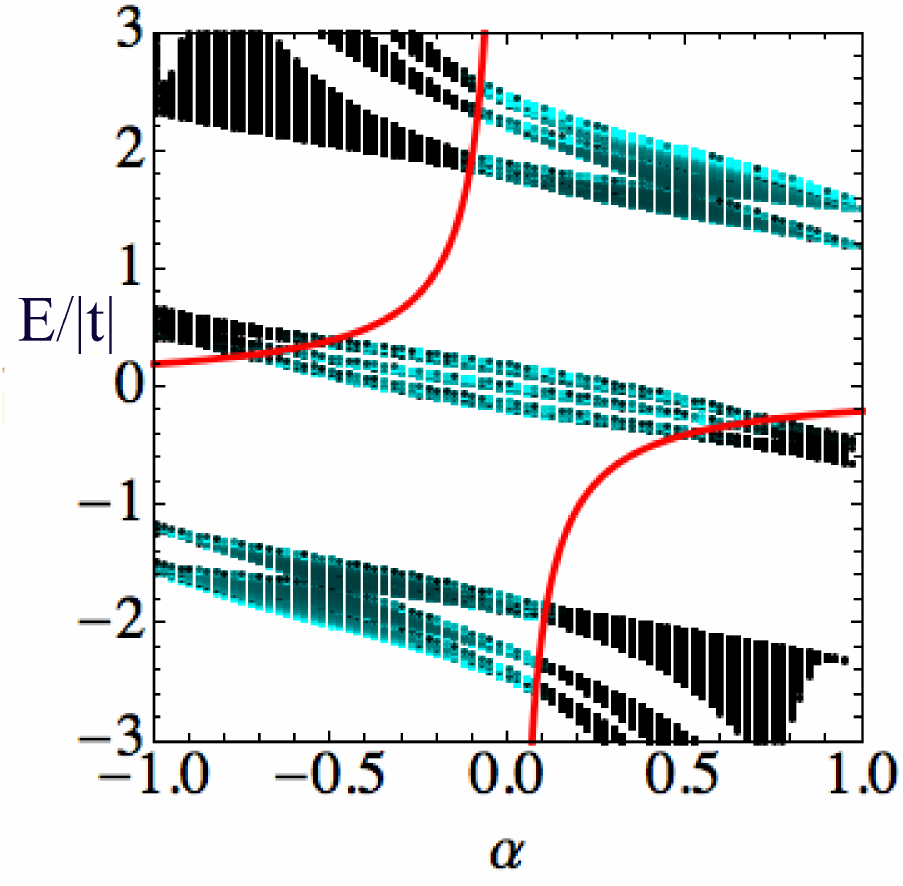}\hspace{0.1cm}\includegraphics[width=4.0cm,height=3.5cm]{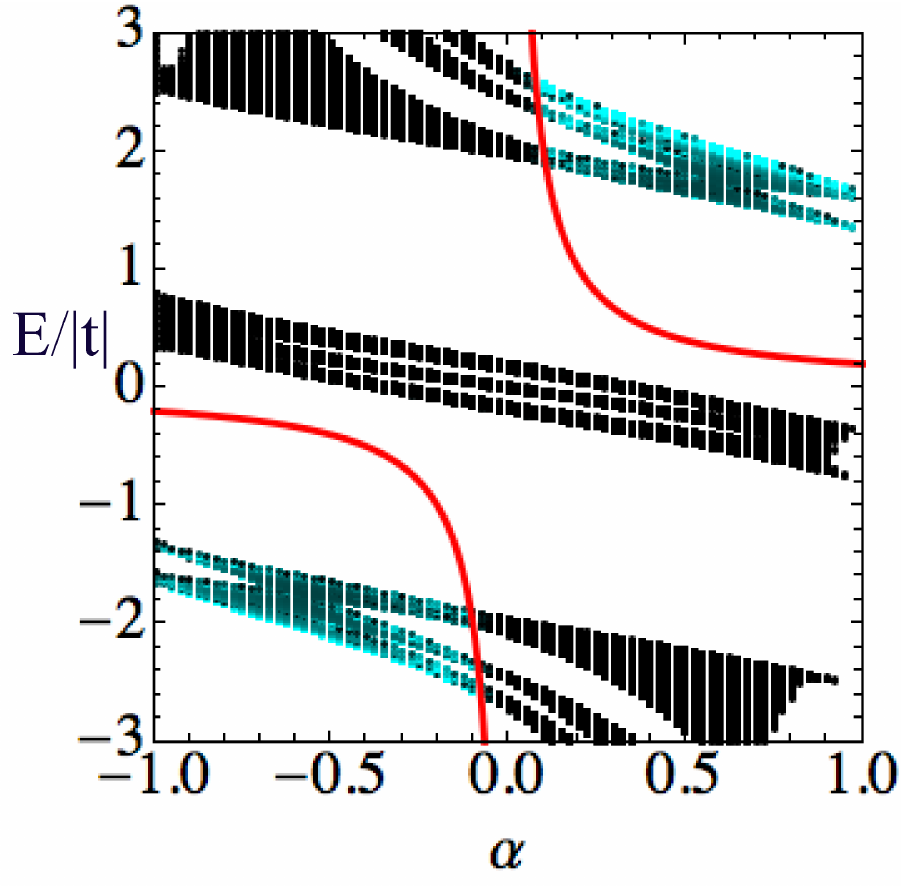}\\
\hspace{0.7cm}\textrm{(c) TDOS}\hspace{4cm}\textrm{(d) TDOS}\\
\vspace{0.1cm}
\includegraphics[width=6.0cm,height=0.8cm]{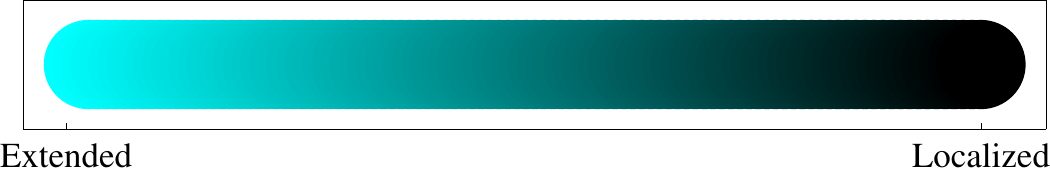}
    \caption{Results for the onsite potential in Eq. (\ref{onsite}).  (a, b): Numerical energy spectrum $E/|t|$ as a function of $\alpha$ for $L=500$ sites tight binding model for $t=-1.0$,  $\lambda/|t|=-0.9$ and $\lambda/|t|=-1.1$. Pure cyan denotes IPR=0 and pure black denotes IPR=1.  (c, d): TDOS plotted after filtering based on the DOS for $L=10,000$ and an expansion order $N_c=32,768$ for  $\lambda/|t|=-0.9$ and $\lambda/|t|=-1.1$. Pure cyan denotes maximum TDOS values between 1 and 10 for extended state and pure black denotes TDOS=0 for localized state. Eigenvalues that have a very small DOS are not plotted, hence the discrepancy in the black regions between the TDOS and the IPR.  Red line is a plot of the analytically obtained critical condition defined in Eq.~(\ref{mainresult}).}  
     \la{ipr_dual1}
\end{figure}
%%%%%%%%%%%%%%%%%%%%%%%%%%%%%%%%%%%%%%%%%%%%%%%%%%%%%%%%%%%%%%%%%%%%%%%%%%%%%

In this letter, we analytically compute the exact mobility edge for a wide class of models specified by Eqs.~(\ref{model},\ref{onsite},\ref{onsite2}). The mobility edge separating the localized and extended states for Eq.~(\ref{onsite}) is given by the following extremely simple closed form expression,
\begin{align}
\alpha E= 2 \sgn(\lambda)(|t|-|\lambda|). \la{mainresult}
\end{align}
This is our central result which we prove by identifying a generalized duality symmetry. Before deriving Eq. (\ref{mainresult}), we show analytical consistency and numerical verification of this condition representing it as a mobility edge. Note that the critical condition must reduce to that of the AAH model for $\alpha=0$. This is indeed the case for $\alpha=0$, as the critical condition [Eq.~(\ref{mainresult})] becomes energy independent giving the familiar self-dual Auby-Andre $|\lambda|=|t|$ critical point. Note that for the AAH model,  the duality transformation is a simple Fourier transformation that maps extended (localized) states in the real space to the localized (extended) states in the Fourier space. This leads to a very special singular continuous Cantor set spectrum at the self-dual point where the critical states can neither be extended nor localized~\cite{AA}. This mapping will not be obvious for the general case and we numerically confirm that this critical condition, defined by Eq.~(\ref{mainresult}), is indeed a localization transition. 

We numerically diagonalize the tight binding model defined in Eq.~(\ref{model}) for $L=500$ sites with periodic boundary conditions. The localization properties of an eigenstate can be numerically quantified using the Inverse Participation Ratio (IPR).
The IPR for an eigenstate $E$ is given as,
\begin{align}
\text{IPR}(E)=\frac{\sum_n|u_n(E)|^4}{(\sum_n |u_n(E)|^2)^2}.
\la{ipr}
\end{align}
For a localized eigenstate,  the IPR approaches the maximum possible value $\sim1$. For an extended state, the IPR is of the order $1/L$, which is vanishingly small in the large system size limit. 

In addition to the IPR,  we define the typical density of states (TDOS) as a new order parameter for the localization transition in quasiperiodic systems. 
\begin{equation}
\rho_t(E)=\exp\left(\frac{1}{L}\sum_{i=1}^L \log \rho_i(E) \right)
\label{eqn:rhot}
\end{equation}
where $\rho_i(E)$ is the local density of states at site $i$ (defined explicitly in the Supplemental Material).  Similar to the Anderson localization in disordered systems~\cite{Janssen-1998, typ, kpm, supp}, this order parameter is finite in the delocalized phase, zero in the localized phase, and goes to zero at the transition.  Due to the quasiperiodic potential, the average over sites resembles a disorder average and we find $\rho_t(E)$ is only equal to the average density of states deep in the delocalized phase~\cite{supp}.  We calculate the TDOS using the kernel polynomial method (KPM), which allows us to reach large system sizes very efficiently (for details see Refs.~\cite{kpm, supp}). For the calculations presented here we consider a chain length of $L=10,000$ and a KPM expansion order $N_c=32,768$.  As the KPM does not determine the energy eigenstates, in order to distinguish between gapped and localized states in the TDOS, we have filtered the energy values plotted in Figs. \ref{ipr_dual1} and \ref{ipr_dual2} based on where the DOS has a finite value~\cite{supp}.  As shown in Figs. \ref{ipr_dual1} and \ref{ipr_dual2}, the results of the TDOS is in excellent agreement with the well known IPR, and thus establish $\rho_t(E)$ as a natural order parameter for the localization transition.  As the TDOS is based on the local density of states, which is related to the imaginary part of the real space Green function, this order parameter can naturally be generalized to interacting many body problems.

We plot the IPR,  the TDOS, and the mobility edge conjectured in Eq.~(\ref{mainresult}) for each energy eigenstate as a function of the dimensionless deformation parameter $\alpha$ for different values of $\lambda$ in Fig.~\ref{ipr_dual1}. %\rem{Both IPR and TDOS of an eigenstate is given by the coloring scheme defined in the legend of the plot. The color scheme is a blend of cyan and black where pure cyan corresponds to the maximally extended state (IPR=0, TDOS~1-10) and pure black denotes the maximally localized state (IPR=1, TDOS=0). The mobility edge conjectured in Eq.~(\ref{mainresult}) is plotted in red. } \jp{[THIS COULD BE REMOVED SINCE IT IS STATED CLEARLY IN THE FIG CAPTION]} 
%We superimpose
Superimposing the IPR and the TDOS calculations with the critical condition conjectured in Eq.~(\ref{mainresult}) we find an excellent agreement between all three.  
%\rem{Fig.~(\ref{ipr_dual1}a, b) is a plot of energy eigenvalue $E/|t|$ as a function of $\alpha$ for fixed values of $\lambda/|t|$. In Figs.~\ref{ipr_dual1}a, c we fix $\lambda/|t|=-0.9$ and $\lambda/|t|=-1.1$ for Figs.~\ref{ipr_dual1}b, d.}
The $\alpha=0$ slice of the plot is the AAH model for which all the states are extended in Figs.~\ref{ipr_dual1}a, c and localized for Figs.~\ref{ipr_dual1}b, d. For $\alpha\ne0$, the states remain extended or localized till it encounters the mobility edge (red line). Across this mobility edge, the IPR and the TDOS go to zero.
%\rem{value discontinuously changes (by two orders of magnitude for 500 sites)} \jp{[I FIND THIS TO BE A POOR CHOICE OF WORDS AS IT SEEMS TO IMPLY ITS A DISCONTINUOUS TRANSITION]}
 indicating a localization transition.
 \begin{figure}[htb!]
  \centering
\includegraphics[width=4cm,height=3.5cm]{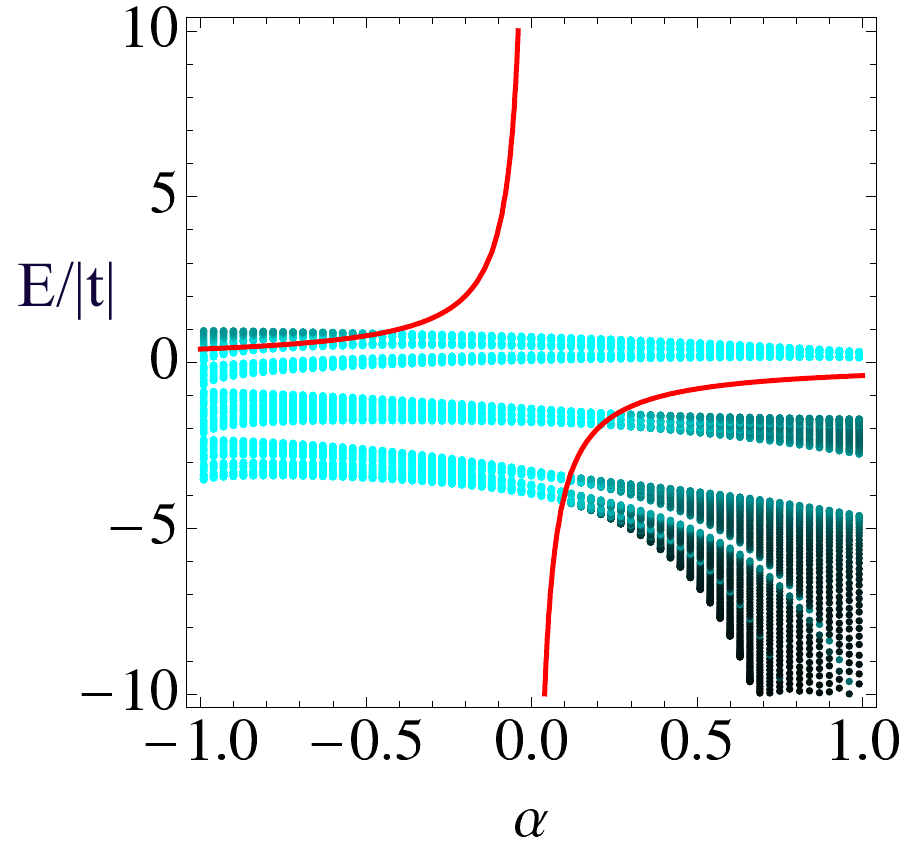}\hspace{0.1cm}\includegraphics[width=4.0cm,height=3.5cm]{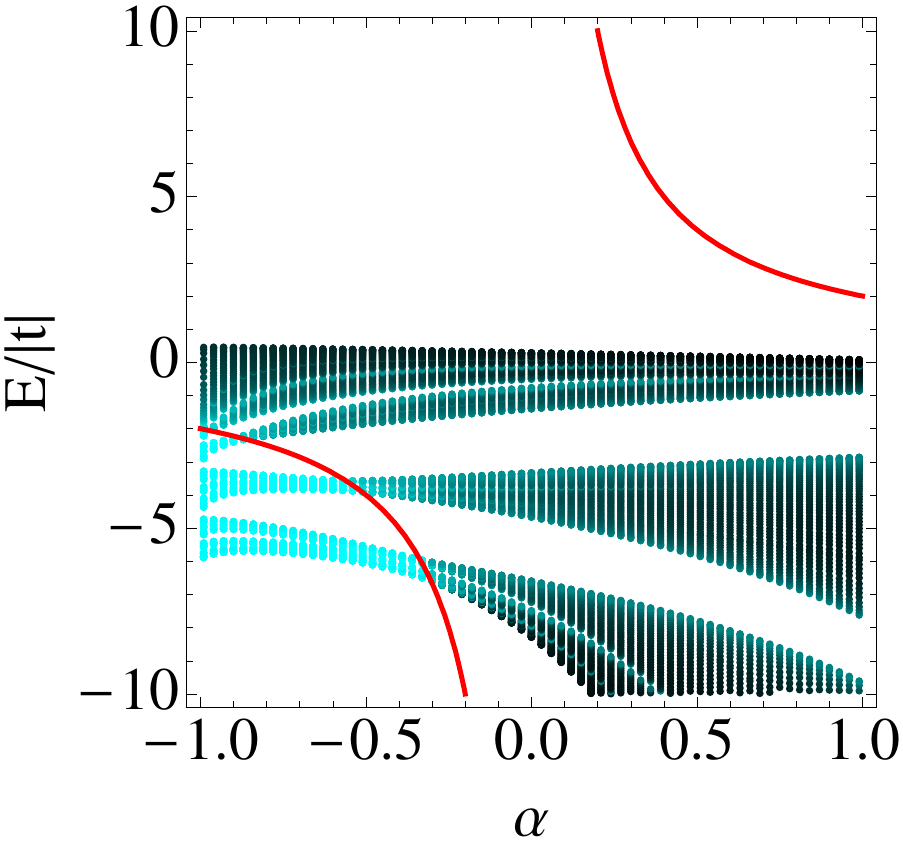}\\
\hspace{0.7cm}\textrm{(a) IPR}\hspace{4cm}\textrm{(b) IPR}\\
\includegraphics[width=4.0cm,height=3.5cm]{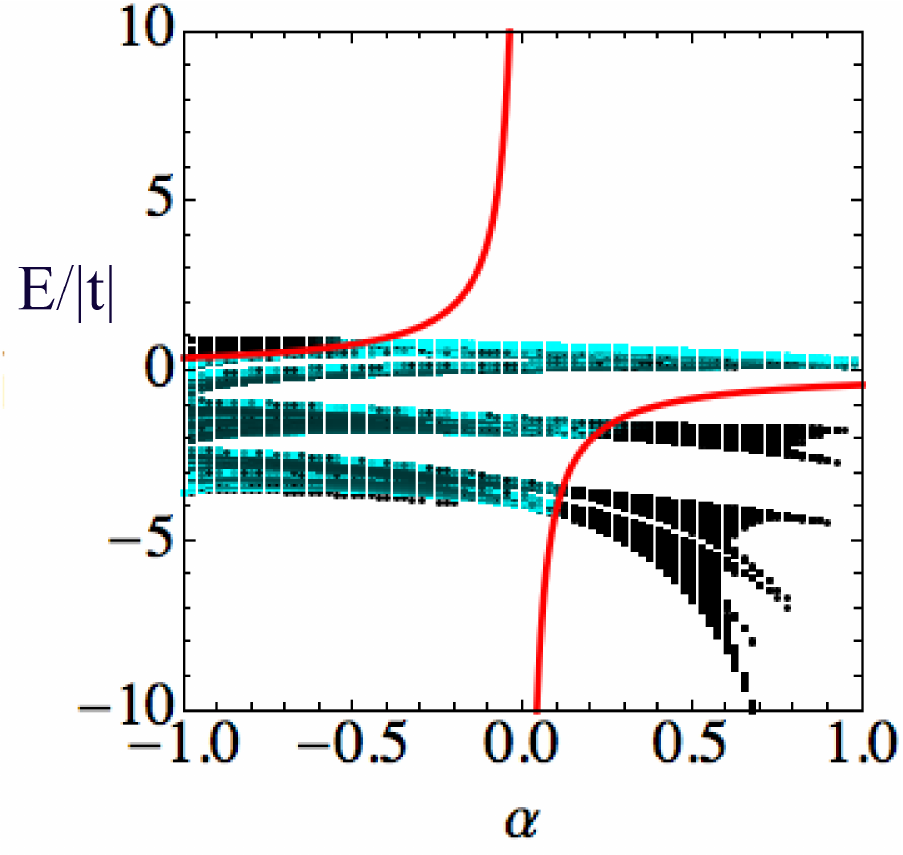}\hspace{0.1cm}\includegraphics[width=4.0cm,height=3.5cm]{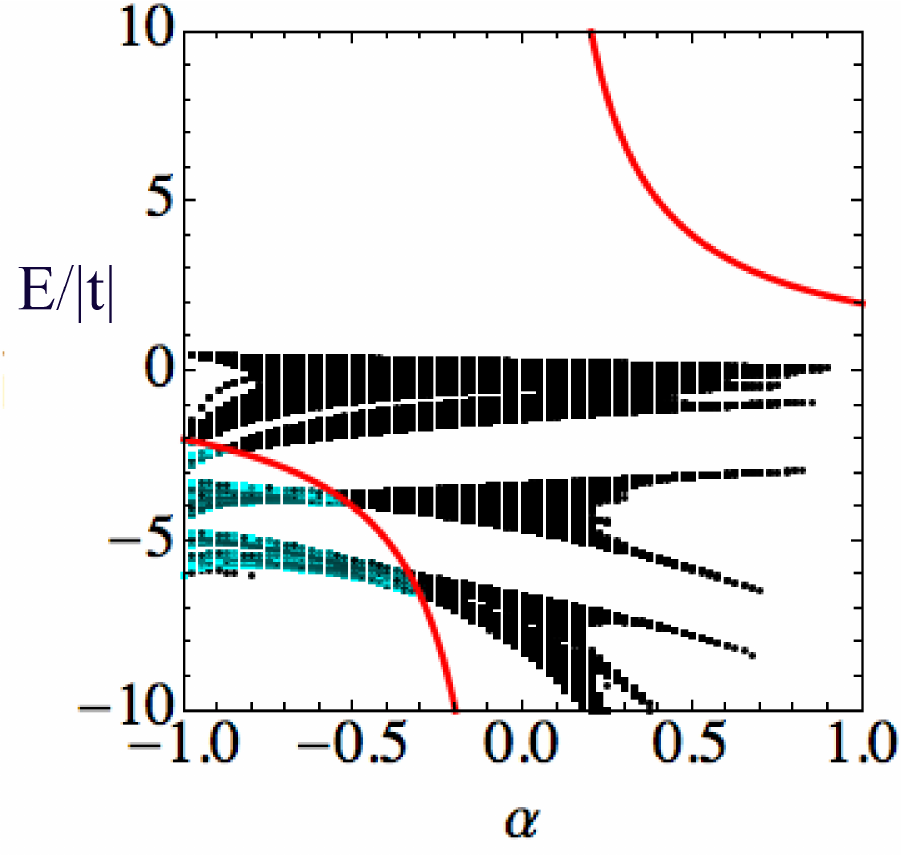}\\
\hspace{0.7cm}\textrm{(c) TDOS}\hspace{4cm}\textrm{(d) TDOS}\\
\vspace{0.1cm}
\includegraphics[width=6.0cm,height=0.8cm]{figlegend.pdf}
    \caption{Results for the onsite potential in Eq. (\ref{onsite2}).  (a, b): Numerical energy spectrum $E/|t|$ as a function of $\alpha$ for $L=500$ sites tight binding model for $t=-1.0$,  $\lambda/|t|=-0.8$ and $\lambda/|t|=-2.0$. Pure cyan denotes IPR=0 and pure black denotes IPR=1.  (c, d): TDOS plotted after filtering based on the DOS for $L=10,000$ and an expansion order $N_c=32,768$ for  $\lambda/|t|=-0.8$ and $\lambda/|t|=-2.0$. Pure cyan denotes maximum TDOS values between 1 and 10 for extended state and pure black denotes TDOS=0 for localized state.  Eigenvalues that have a very small DOS are not plotted, hence the discrepancy in the black regions between the TDOS and the IPR. Red line is a plot of the analytically obtained critical condition defined in Eq.~(\ref{mainresult}).}  
  \la{ipr_dual2}
\end{figure}

In the following, we consider a different family of nearest neighbor models that satisfies the same critical condition defined in Eq.~(\ref{mainresult}). This family is defined by the following onsite potential, 
%\bea
%V_{n}(\alpha,\phi)&=&2\lambda\frac{1-\cos(2\pi n b+\phi)}{1+\alpha\cos(2\pi n b+\phi)}\nonumber\\
%&=&2\lambda\times\begin{cases}
%1& \text{as}\ \alpha\rightarrow -1\\
%\\
%1-\cos(2\pi n b+\phi) & \text{as}\ \alpha\rightarrow 0\\
%\\
%\tan^2(\frac{2\pi n b+\phi}{2})& \text{as} \ \alpha \rightarrow 1
%\end{cases}
%\la{onsite2}
%\eea
\begin{equation}
V_{n}(\alpha,\phi)=2\lambda\frac{1-\cos(2\pi n b+\phi)}{1+\alpha\cos(2\pi n b+\phi)}.
\la{onsite2}
\end{equation}
Note that $V_{n}(\alpha,\phi)$ in Eq.~(\ref{onsite2}) connects different tight binding models compared to the potential in Eq.~(\ref{onsite}). The limiting cases of $V_{n}(\alpha,\phi)$ are as follows; for $\alpha=-1$ it corresponds to a constant onsite energy $2\lambda$, for $\alpha=0$ it corresponds to a rescaled AAH model, and for $\alpha=1$ it corresponds to the closed form singular potential given by $\tan^2\frac{2\pi n b+\phi}{2}$.  In Fig.~(\ref{ipr_dual2}), we plot the numerical spectrum as a function of $\alpha$ with color coded IPR (Fig.~\ref{ipr_dual2}a, b) and TDOS (Fig.~\ref{ipr_dual2}c, d) for $\lambda/|t|=-0.8, \  -2.0$. The critical condition (shown in red line ) is in excellent agreement with the separation of the localized and extended states as indicated by the IPR and TDOS. We now emphasize a special feature of this 1D model which manifests the non-perturbative nature of the 3D Anderson localization. Note that $\alpha=-1$ gives a disorder free constant potential and all the states must be extended. Consequentially, all the states lie just below the critical line in the extended regime for all values of $\lambda/|t|$. Infinitesimal deviation from the $\alpha=-1$ point manifests localized states across this critical line. The number of available localized states depends on the value of $\lambda/|t|$. This feature, showing up clearly in our 1D incommensurate model, is one of the striking manifestations of Anderson localization phenomenon, where any infinitesimal disorder completely localizes some eigenstates forming a sharp mobility edge defined by Eq.~(\ref{mainresult}). 
%%%%%%%%%%%%%%%%%%%%%%%%%%%%%%%%%%%%%%%%%%%%%%%%%%%%%%%%%%%%%%%%%%%%%%%%%%%%%

%%%%%%%%%%%%%%%%%%%%%%%%%%%%%%%%%%%%%%%%%%%%%%%%%%%%%%%%%%%%%%%%%%%%%%%%%%%%%%%%%%%%%%%%

 %
 %%%%%%%%%%%%%%%%%%%%%%%%%%%%%%%%%%%%%%%%%%%%%%%%%%%%%%%%%%%%%%%%%%%%%%%%%%%%%%%%%%%%%%%%
%\begin{figure}[htb!]
 % \centering
%\textrm{(a)}\hspace{4cm}\textrm{(b)}
%\includegraphics[width=4cm,height=3cm]{fig0a.pdf}\hspace{0.1cm}\includegraphics[width=4.0cm,height=3cm]{fig0b.pdf}\\
 % \caption{}
 % \label{exponent}
%\end{figure}
%%%%%%%%%%%%%%%%%%%%%%%%%%%%%%%%%%%%%%%%%%%%%%%%%%%%%%%%%%%%%%%%%%%%%%%%%%%%%%%%%%%%%%%%
\emph{Self-duality:} Having presented compelling numerical evidence that the condition conjectured in Eq.~(\ref{mainresult}) is a critical point of a localization transition, we now analytically derive this condition. We rewrite the model defined in Eq.~(\ref{model}) for the onsite potentials defined in Eqs.~(\ref{onsite}, \ref{onsite2}) in the following form,
\begin{align}
t(u_{p-1}+u_{p+1})+g \chi_p(\beta) u_p=(E+2\lambda \cosh \beta)u_p.
\label{model1}
\end{align} 
Where we have defined the onsite potential $\chi_p(\beta)$ as 
\begin{align}
\chi_p(\beta)=\frac{ \sinh \beta }{\cosh\beta-\cos (2\pi p b+\phi)}=\sum^{\infty}_{r=-\infty} e^{-\beta |r|} e^{i r( 2\pi p b+\phi)}\la{id},
\end{align}
with $1/\alpha=\cosh \beta$ for $\alpha>0$, $t>0$, and $\lambda>0$. We can absorb the sign of $\alpha$ and $t$ as a redundant phase shift in the cosine term. Note that by definition, a change in the sign of $\lambda$ can be absorbed as a phase only in combination with the change in sign of $\alpha$. We restore all the sign dependence of the duality condition at the end of our calculation. The parameter $g$ is model dependent and is given by $g=2 \lambda\cosh \beta/\tanh \beta$ for the onsite potential defined in Eq.~(\ref{onsite}) and $g=2\lambda(1+\cosh \beta)/\sinh \beta$ for the onsite potential in Eq.~(\ref{onsite2}) .  The above parametrization will help us to identify the hidden duality symmetry of this model. Note that Eq.~(\ref{model1}) can be deformed without breaking the duality symmetry by making an arbitrary choice for the parameter $g\equiv g(\alpha,\lambda)$ and $E\equiv E(\alpha, \lambda, t)$ to design several onsite potentials with exact mobility edges. In the following we can set the overall phase $\phi=0$ without loss of generality. Now we define the following ansatz for the duality transformation under which the model in Eq.~(\ref{model1}) is self dual 
\begin{align}
f_k=\sum_{m n p}e^{i 2\pi b(k m+m n +n p) }\chi_n^{-1}(\beta_0) u_p \la{trans}.
\end{align}
We define $\sum_n \equiv \sum^{\infty}_{n=-\infty}$ from hereon. We have defined $\beta_0$ below Eq.~(\ref{model2}). 
The above transformation can be viewed as three independent transformations acting on $u_p$.
In the following, we show how these three transformations act on the tight binding model in Eq.~(\ref{model1}) resulting
in the final tight binding model for $f_k$. We multiply Eq.~(\ref{model1}) by  $e^{i 2\pi b n p}$ and performing a summation 
over $p$, we obtain,  
\begin{align}
\omega \chi^{-1}_n(\beta_0) v_n=g \sum_r e^{-\beta |n-r|} v_r.
\la{model2}
\end{align}  
%\begin{align}
%2t \cos (2\pi n b)v_n+g \sum_p e^{i2\pi b n p} \chi_p u_p=(E+2\lambda \cosh \beta)v_n.
%\end{align} 
Here we have defined $v_n=\sum_p e^{i 2\pi b n p} u_p$, $2t\cosh \beta_0=E+2\lambda \cosh \beta$ and $\omega=2t\sinh \beta_0$. Now we multiply Eq.~(\ref{model2}) by $e^{i 2\pi b m n}$ and carry out summation over 
$n$. The resulting equation can be rewritten in terms of $w_m=\sum_n e^{im(2\pi b  n)}\chi^{-1}_n(\beta_0)v_n$ as,
 \begin{align}
\omega \chi^{-1}_m(\beta)w_m=g \sum_{r } e^{-\beta_0 |m-r|} w_r. \la{model3}
\end{align}
Note that the above step is a generalized transformation encoded in the definition of $w_m$. This is the final operation of our generalized transformation where we define $f_k=\sum_m e^{i2\pi b m k}w_m$. In the final step, we multiply Eq.~(\ref{model3}) by $e^{i2\pi b m k}$ and sum over $m$ to obtain 
the following tight binding model  for $f_k$,
   \begin{align}
t(f_{k+1}+f_{k-1})+g\frac{\sinh \beta}{\sinh \beta_0}\chi_k(\beta_0) f_k=2t\cosh \beta f_k. \la{dualmodel}
\end{align}
We have derived the result we set out to prove. The above tight binding model in terms of $f_k$ in Eq.~(\ref{dualmodel}) is explicitly self dual 
to the original tight binding model defined in Eq.~(\ref{model1}) if $\beta_0=\beta$. This self duality condition can be expressed in terms of $E$, $\alpha$, $t$ and $\lambda$,
% \begin{align}
%\alpha E= 2(t-\lambda)
% \end{align}
restoring the sign dependence of the duality condition, we obtain our main result
 \begin{align}
 \alpha E= 2\sgn(\lambda)(|t|-|\lambda|)
\la{dualitycond}
\end{align}
which holds for $\alpha \in (-1, 1)$ and $\forall \, \lambda, t$. Thus we have proved the condition for self duality which we proposed and numerically verified as defining the mobility edge of a localization transition. 

\emph{Experimental design:-} Quasiperiodic 1D lattices have been realized in ultra cold atoms (Bose-Einstein condensate (BEC) of $K^{39}$ atoms) by a standing wave arrangement of two laser beams with mutually incommensurate wave vector~\cite{roati08}.  The quasiperiodic potentials considered in this letter can be systematically engineered by a controlled application of a series of standing wave laser beams. The experimental schematic becomes transparent by considering the Cosine Fourier series of the onsite potential defined in Eqs.~(\ref{onsite} and \ref{onsite2}),
\begin{align}
 V_n(\beta, \phi)=\frac{a_0(\beta)}{2}+\sum_{r=1}^{\infty}a_r(\beta) \cos (r (2\pi n b+\phi)),
 \la{expand}
\end{align}
where the coefficients are given by $a_0=-4\lambda \cosh\beta$, $a_r=2g e^{-\beta r}$.  Note that $g$ is an overall model dependent constant defined in Eq.~(\ref{model1}). The $r=1$ term is the AAH term which has been realized by a standing wave laser of wave vector $k_1$ superimposed on the underlying lattice potential generated by another standing wave laser beam of wave vector $k_2$~\cite{roati08, modugno}. The incommensuration parameter is defined by the wave vector ratio $b=k_1/k_2$.  The $r>1$ terms can be realized by simply adding a series of standing wave laser beams with a wave vector that is an integer multiple of  $k_1$. The intensity of the $r^{th}$ harmonic laser beam is determined by the coefficient $a_r$. For a small value of $\alpha$ (large $\beta$), the Fourier series can be truncated with few Fourier components. The mobility edge is extremely pronounced even for small values of $\alpha$ if we fine tune the model parameters to the critical AAH model ($|\lambda|=|t|$). As shown in Figs.~(\ref{ipr_dual1}b and \ref{ipr_dual2}b), even an infinitesimal $\alpha$ manifests a sharp mobility edge.  Experiments should trace both static (momentum distribution of a stationary state) and dynamical  (diffusion dynamics) properties of the localized BEC condensate to demonstrate a clear localization transition. In addition to ultracold atoms, the localization of the AAH model was also observed in quasiperiodic photonic waveguides.  For this setup, the localization is quantified by directly monitoring the IPR of the injected wave packet~\cite{lahini2009, krausfib}. %Recent photon pumping experiments with complex quasiperiodic models~\cite{krausfib} demonstrate the capability for design flexibility towards realizing onsite potentials considered in this work.

\emph{Conclusion:-} In this work we have analytically demonstrated the surprising existence of a mobility edge in a wide class of 1D nearest neighbor tight binding models with quasiperiodic onsite potentials. We show that the analytical critical condition is in excellent agreement with the localization properties obtained from the numerical computation of IPR and TDOS. We outlined a concrete design schematic and observation methodology of our results within existing experimental setups.

\emph{Acknowledgements}: SG and JHP would like to thank Xiaopeng Li for useful discussions. This work is supported by JQI-NSF-PFC. The authors acknowledge the University of Maryland supercomputing resources (http://www.it.umd.edu/hpcc) made available in conducting the research reported in this paper.

\bibliographystyle{my-refs}

\bibliography{references.bib}

\onecolumngrid
\newpage 
\appendix 

\section{Typical density of states}
In the supplemental material we give the details of the kernel polynomial method (KPM) we used to calculate the local density of states, for more details see Ref.~\cite{kpm}.  The local density of states at site $i$ is defined as
\begin{equation}
\rho_i(E) = \sum_k |\langle k| i \rangle|^2\delta(E-E_k)
\end{equation}
where $| k \rangle$ denotes an eigenstate in real space and $E_k$ its corresponding eigenvalue.  From this we can also determine the average density of states 
\begin{equation}
\rho_a(E) = \frac{1}{L}\sum_i \rho_i(E).
\label{eqn:rhoa}
\end{equation}
 Central to the KPM, we expand the local density of states in terms of Chebyshev polynomials $T_n(x)$ and truncate the expansion at order $N_c$. As Chebyshev polynomials are only defined on an interval $[-1,1]$ we rescale the Hamiltonian so that its eigenvalues fall within the corresponding range, which can be done from a simple transformation $H = aH' + b$ where $a=(E_{max}-E_{min})/(2-\epsilon)$ and $b=(E_{max}+E_{min})/2$ are related to the maximum and minimum eigenvalues and we take $\epsilon=0.01$ to ensure the eigenvalue spectrum does not exceed the bounds. We estimate $E_{max}$ and $E_{min}$ efficiently using the Lanczos method. In short, the KPM reduces the problem of diagonalizing the Hamiltonian into calculating the coefficients of the expansion, which yields~\cite{kpm} (for site $i$)
 \begin{equation}
  \mu_n(i) = \langle i | T_n(H') | i\rangle
  \end{equation}
  where $T_n(H')$ is a Chebyshev polynomial of the rescaled Hamiltonian matrix and
  this can be done using only matrix-vector multiplications. As the tight binding model with an on-site potential is sufficiently sparse, the calculation can be done very efficiently allowing us to study sufficiently large system sizes and keeping a very large number of Chebyshev moments. Well known from Fourier series, truncating the expansion can lead to artificial oscillations (Gibbs oscillations) in the calculation, to avoid this we use the Jackson Kernel to filter these effects out~\cite{kpm}. 
    %%%%% Figure alpha=0
\begin{figure}[htb!]
\centering
\begin{minipage}{0.4\textwidth}
\centering
  \includegraphics[angle=-90,width=7cm]{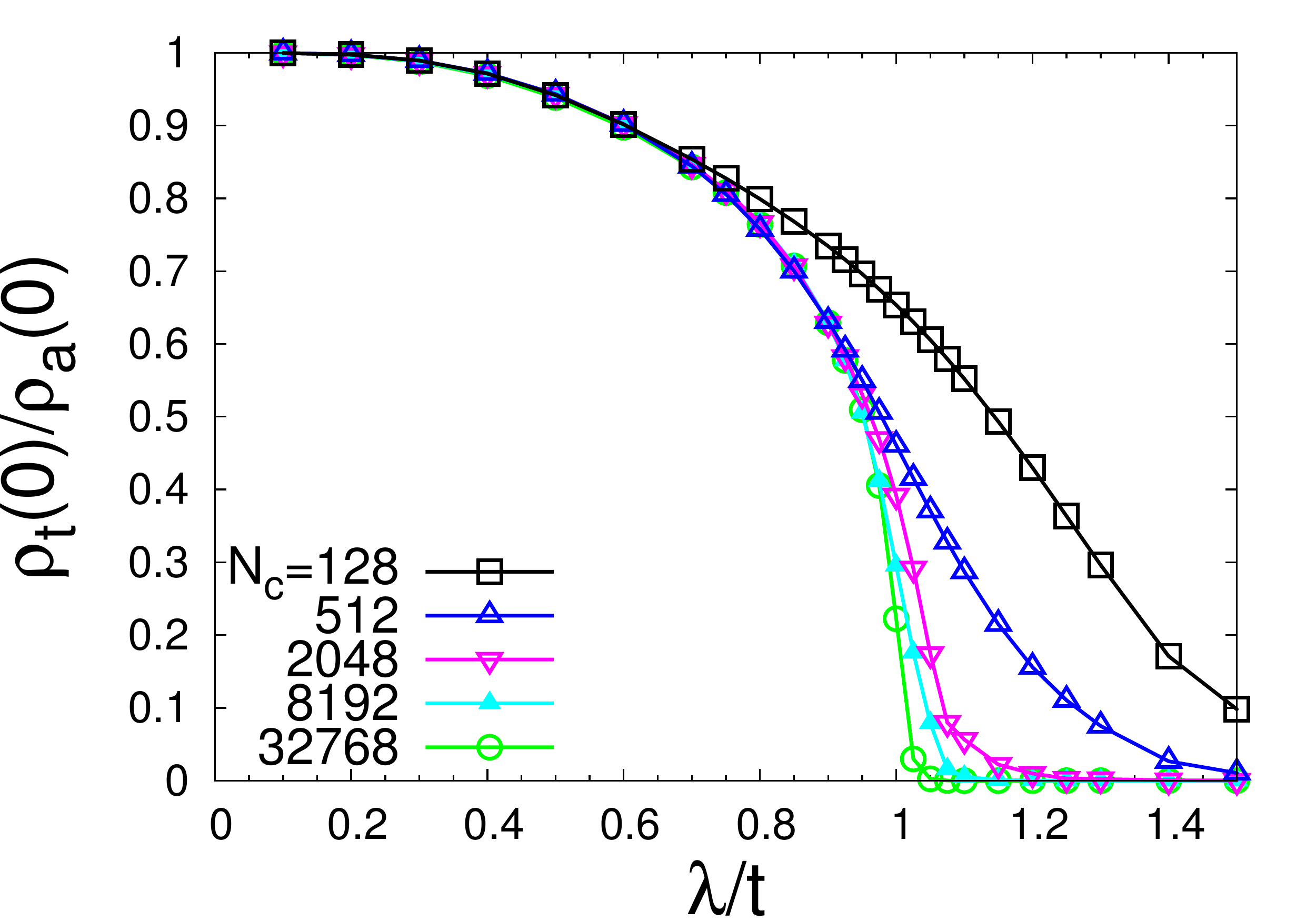}
  \end{minipage}
  \begin{minipage}{0.4\textwidth}
\centering
  \includegraphics[angle=-90,width=7cm]{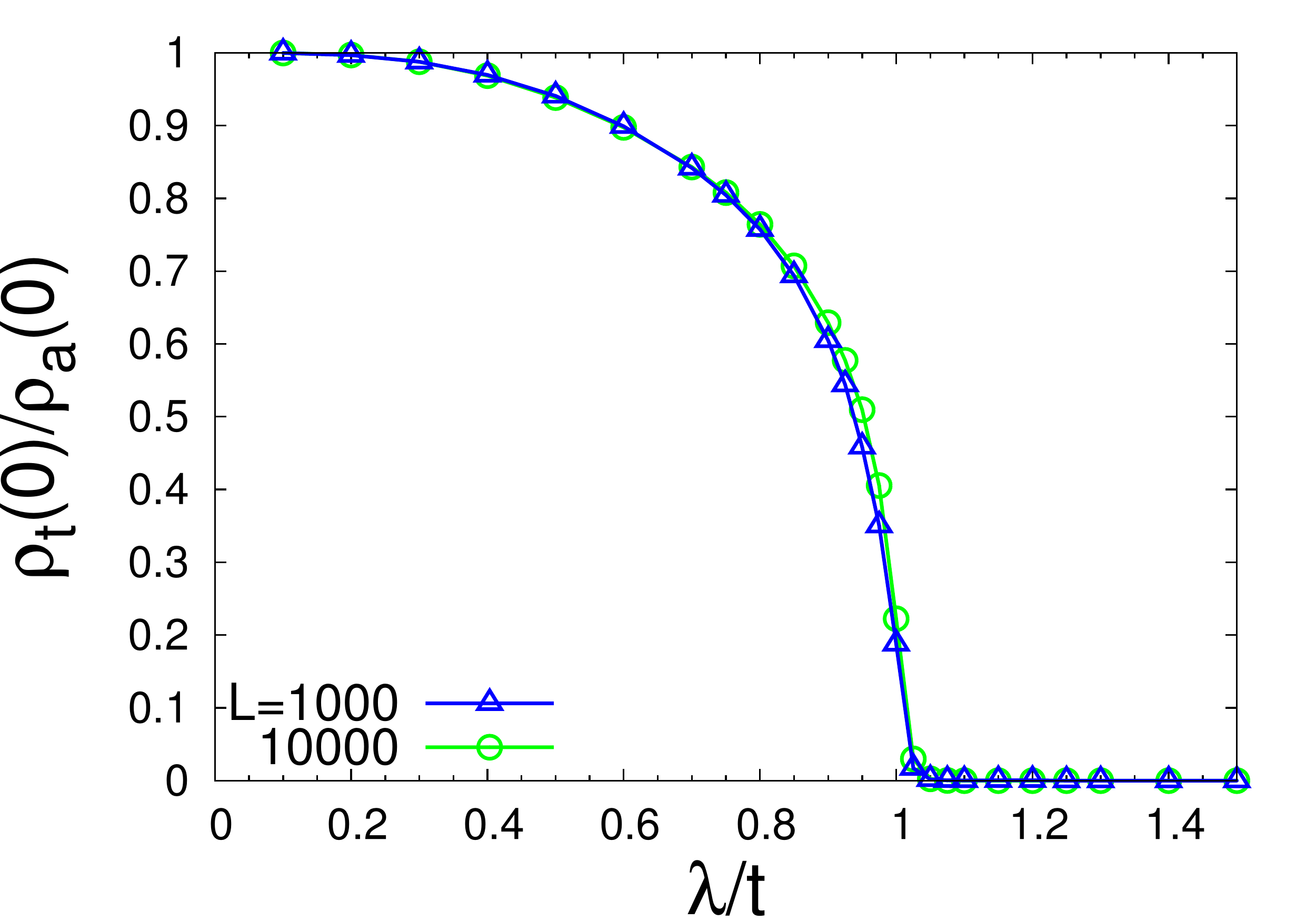}
  \end{minipage}
\caption{Typical density of states normalized by the average density of states in the center of the band ($E=0$) as a function of $\lambda/t$ for the AAH model ($\alpha=0$). (Left) for various different Chebyshev expansion orders $N_c$ and a system size of $L=10,000$ sites.  We find increasing the order of the expansion leads to a clear localization transition at the known value of $\lambda_c/t=1$. (Right) Typical density of states for system sizes $L=1,000$ and $10,000$ for $N_c=32,768$, we find the transition becomes sharper as we increase the length of the chain, but has more or less converged after reaching $L=10,000$.  Deep within the delocalized phase, the typical and average density of states are equal, where as across the localization transition the average density of states remains finite while the typical density of states goes to zero.}
\label{fig:rhot_a=0}
\end{figure}
%%%%% Figure alpha=0  
 
  For all of the calculations presented in the main text we have considered a system size of $L=10,000$ and calculated $N_c = 32,768$ moments such that the typical density of states (defined in Eq. (5) of main text) is smooth. As shown in Fig.~\ref{fig:rhot_a=0}, focusing on the center of the band at $E=0$ in the AAH model  ($\alpha=0$), the localization transition at $\lambda_c/t=1$ is reproduced quite clearly.
  
\begin{figure}[htb!]
\centering
\begin{minipage}{0.4\textwidth}
\centering
  \includegraphics[width=6cm]{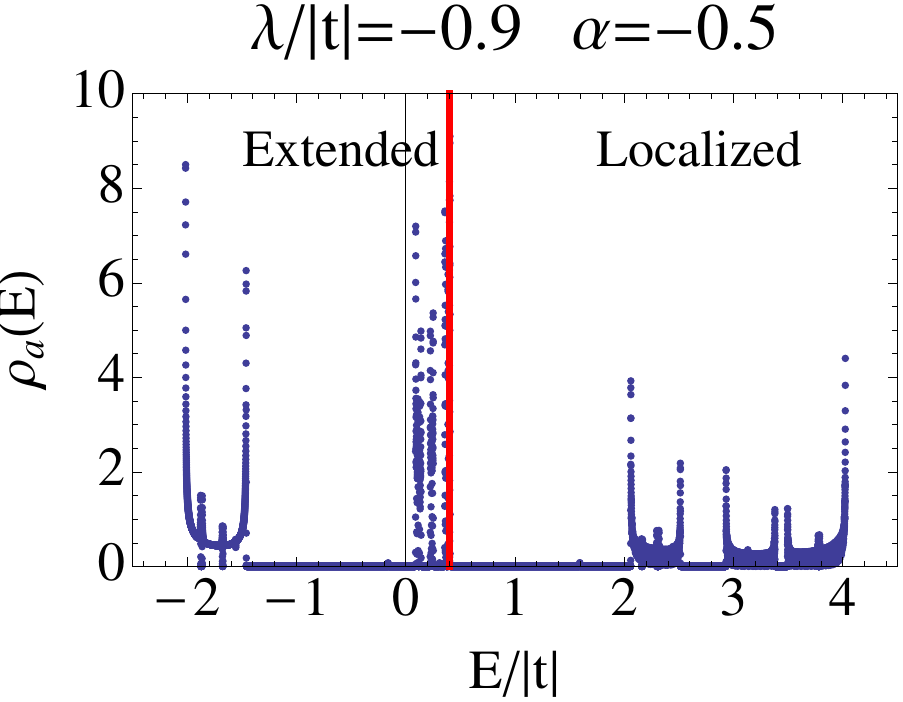}
  \end{minipage}
  \begin{minipage}{0.4\textwidth}
\centering
  \includegraphics[width=6cm]{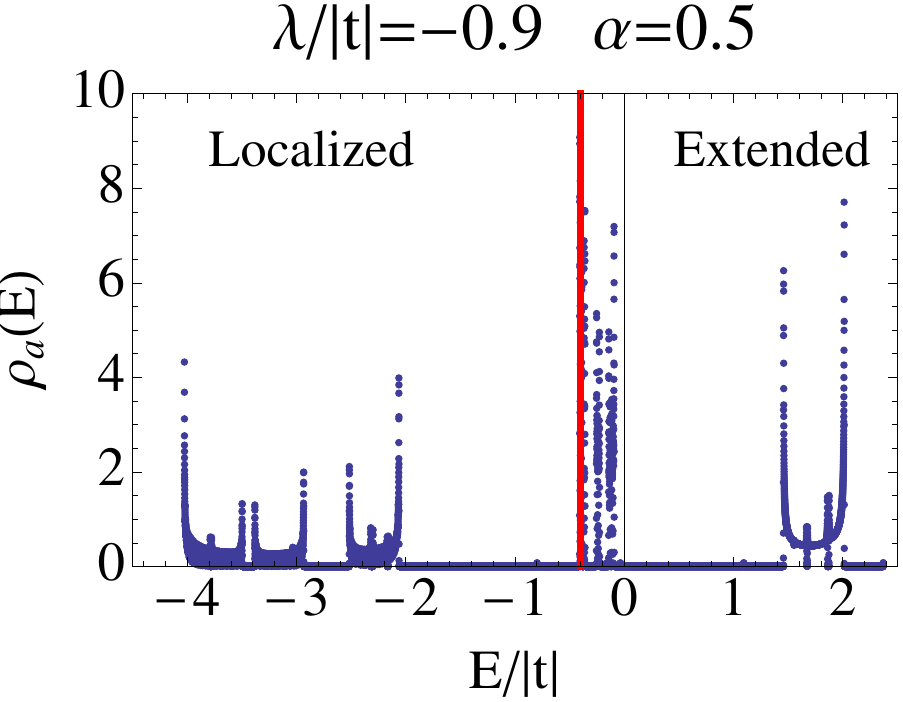}
  \end{minipage}
  \begin{minipage}{0.4\textwidth}
\centering
  \includegraphics[width=6cm]{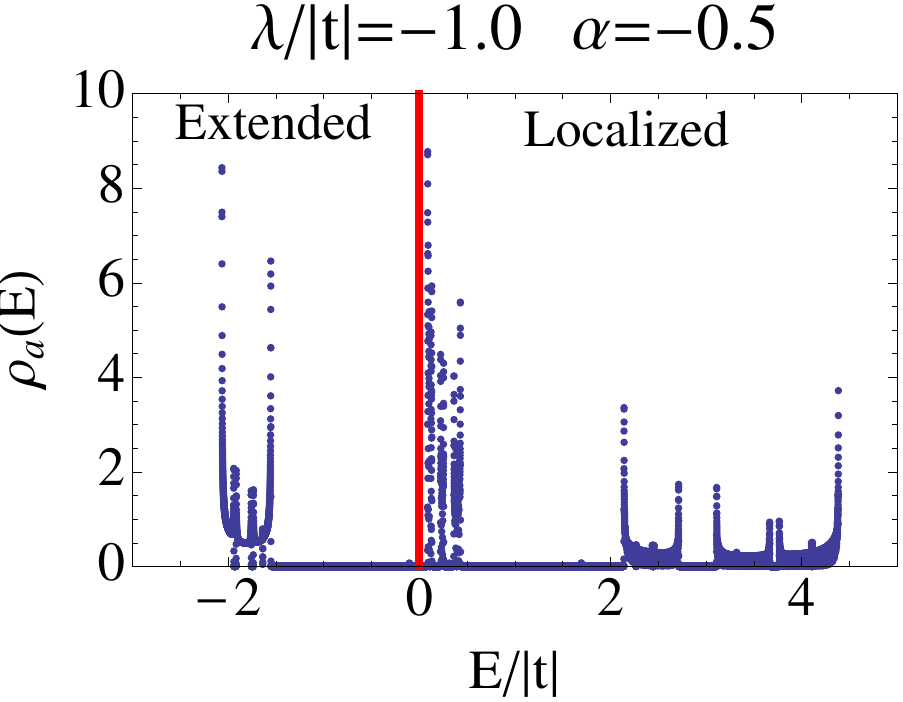}
  \end{minipage}
  \begin{minipage}{0.4\textwidth}
\centering
  \includegraphics[width=6cm]{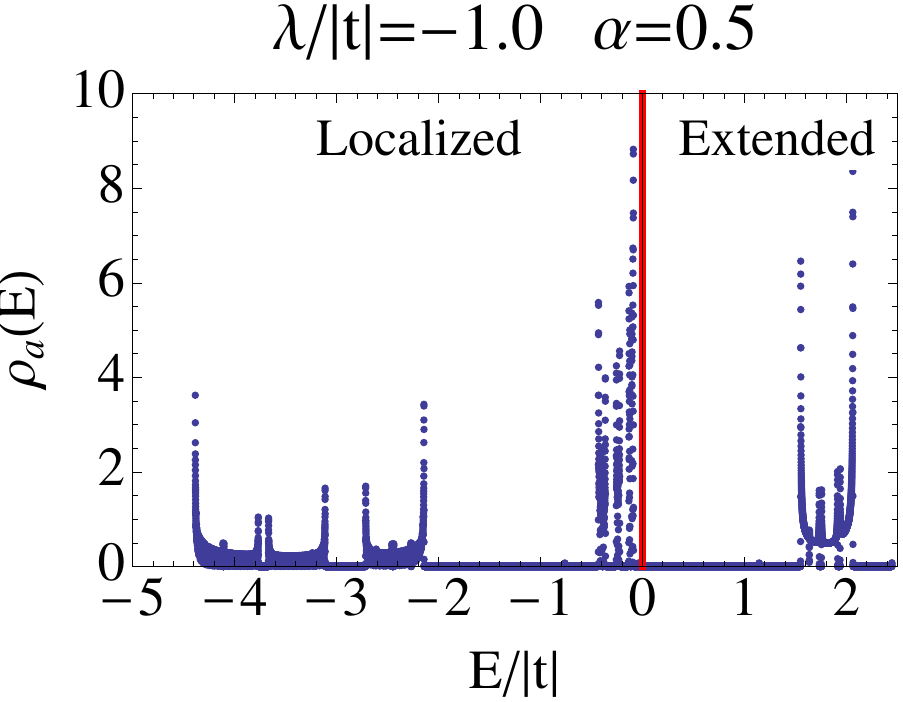}
  \end{minipage}
  \begin{minipage}{0.4\textwidth}
\centering
  \includegraphics[width=6cm]{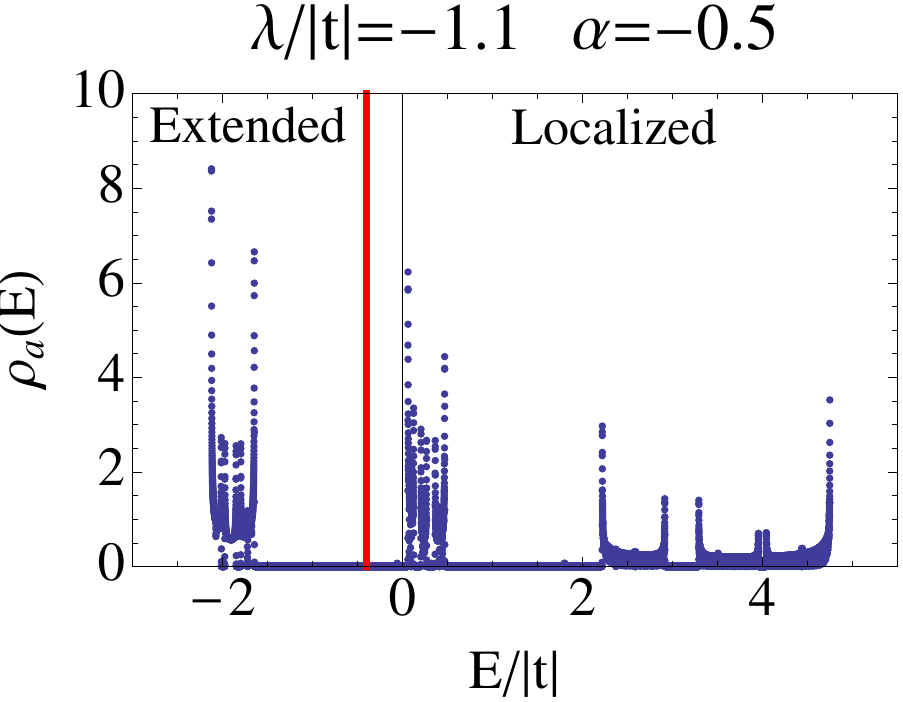}
  \end{minipage}
  \begin{minipage}{0.4\textwidth}
\centering
  \includegraphics[width=6cm]{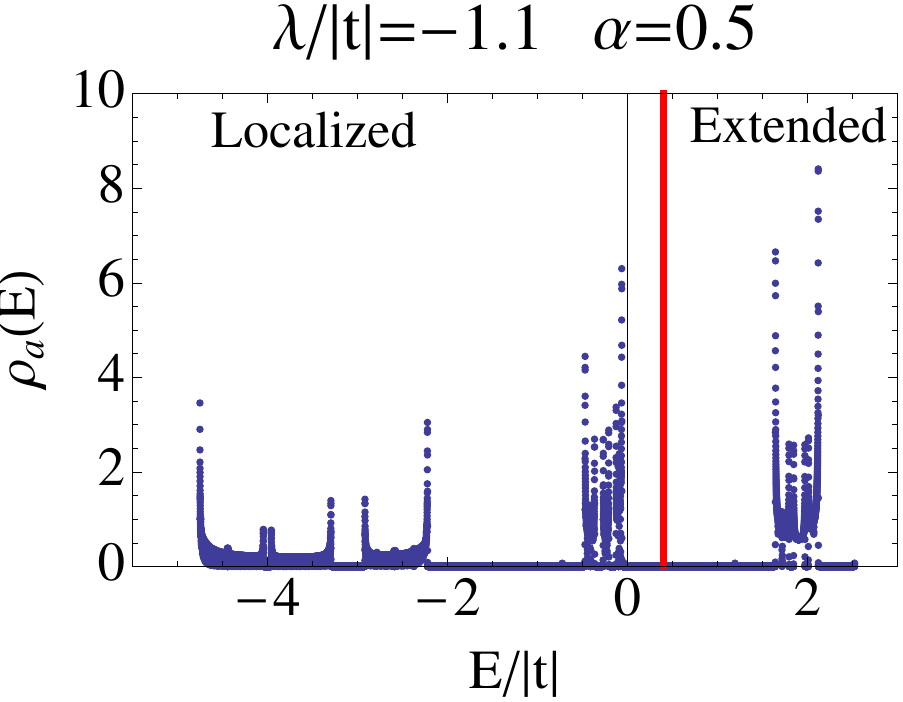}
  \end{minipage}
   \begin{minipage}{0.4\textwidth}
\centering
  \includegraphics[width=6cm]{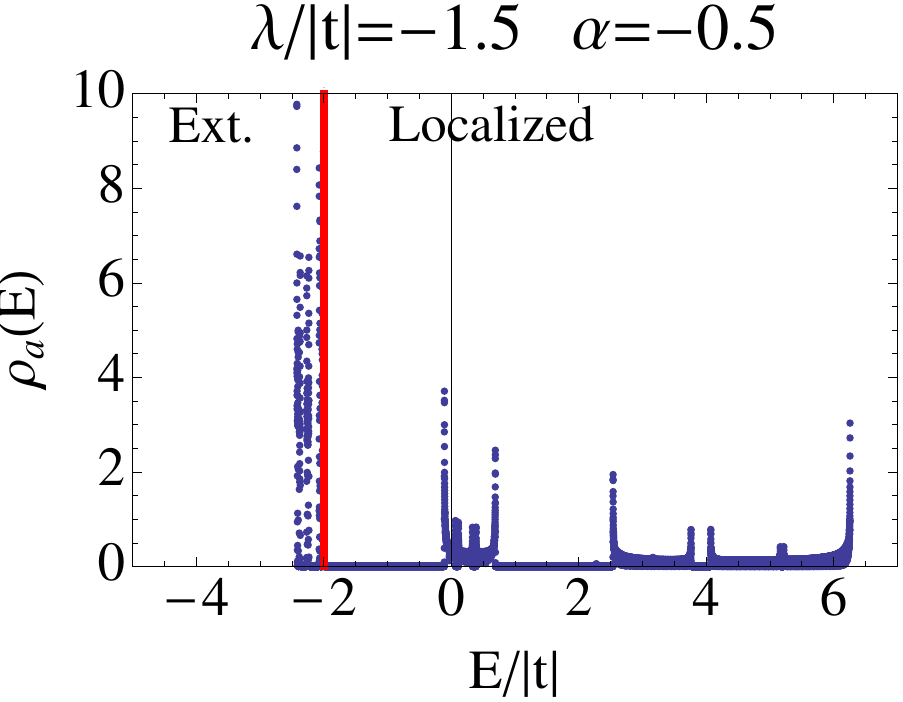}
  \end{minipage}
  \begin{minipage}{0.4\textwidth}
\centering
  \includegraphics[width=6cm]{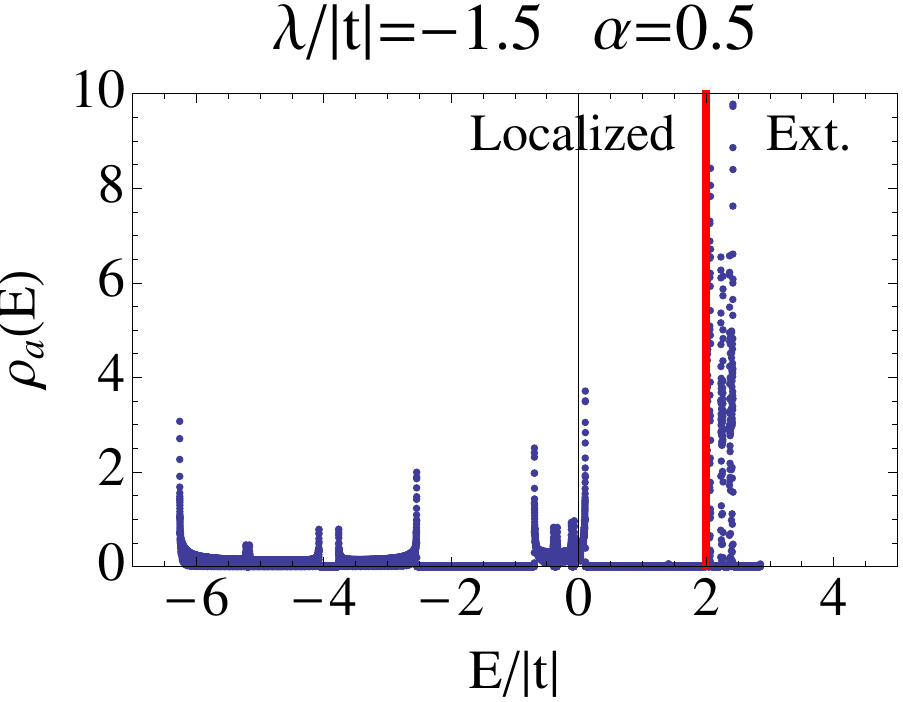}
  \end{minipage}
\caption{Density of states as a function of energy $E$ for the potential defined in Eq. (\ref{onsite}) for various values of $\lambda$ and $\alpha$ as specified within each figure.  We also show the analytic result for the location of the mobility edge in red.}
\label{fig:rhot_a1}
\end{figure}

\begin{figure}
\centering
\begin{minipage}{0.3\textwidth}
\centering
  \includegraphics[width=4.5cm]{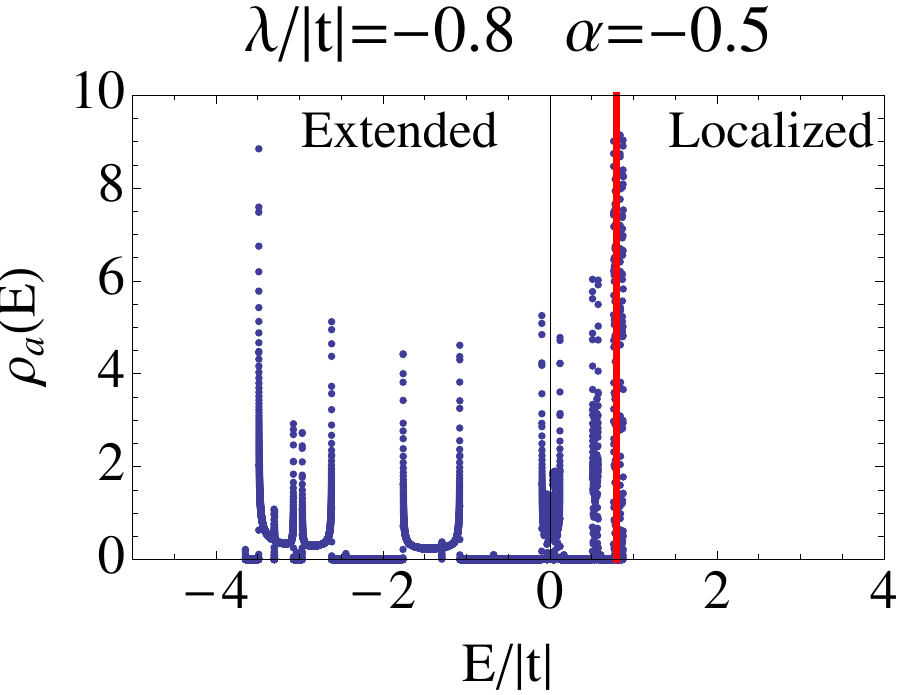}
  \end{minipage}
  \begin{minipage}{0.3\textwidth}
\centering
  \includegraphics[width=4.5cm]{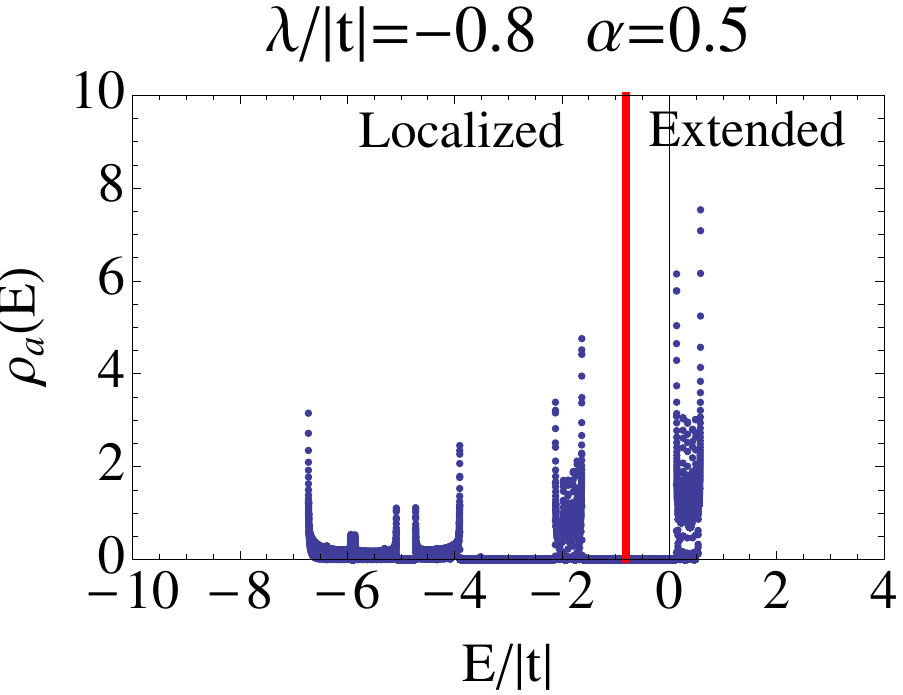}
  \end{minipage}
  \begin{minipage}{0.3\textwidth}
\centering
  \includegraphics[width=4.5cm]{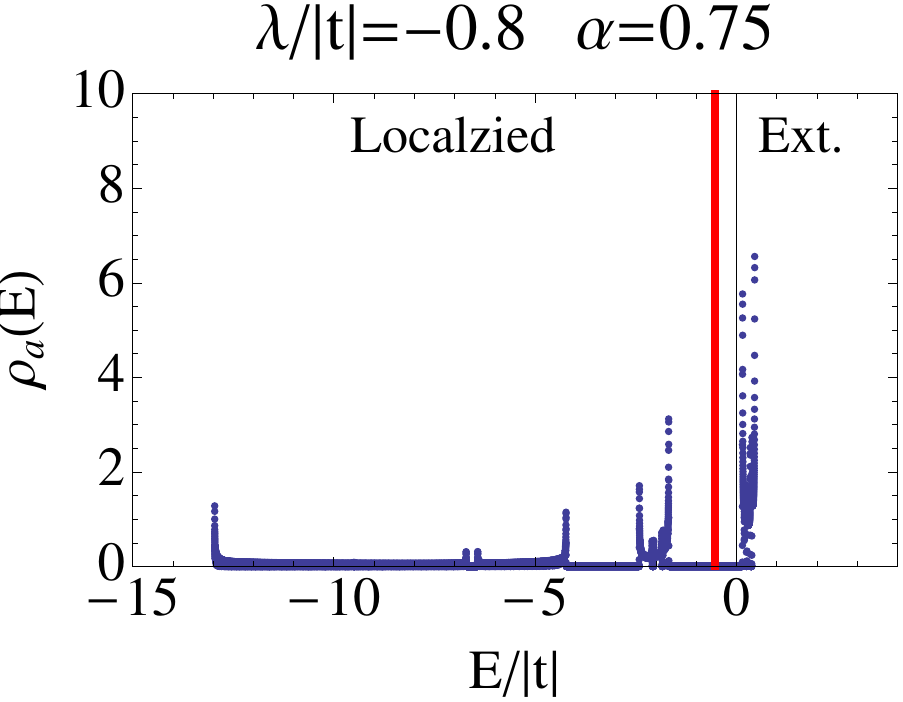}
  \end{minipage}
  \begin{minipage}{0.3\textwidth}
\centering
  \includegraphics[width=4.5cm]{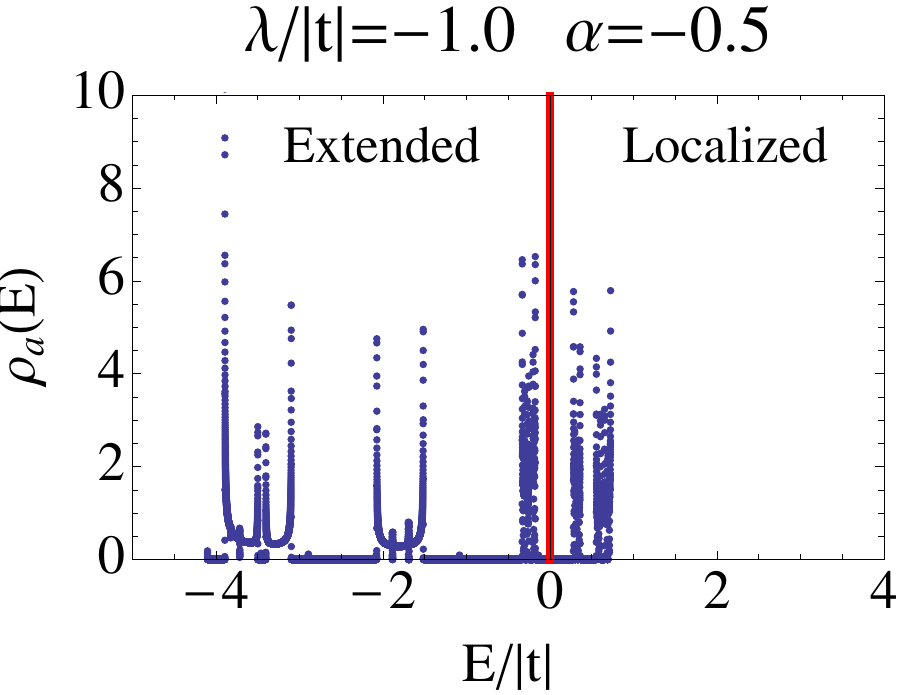}
  \end{minipage}
  \begin{minipage}{0.3\textwidth}
\centering
  \includegraphics[width=4.5cm]{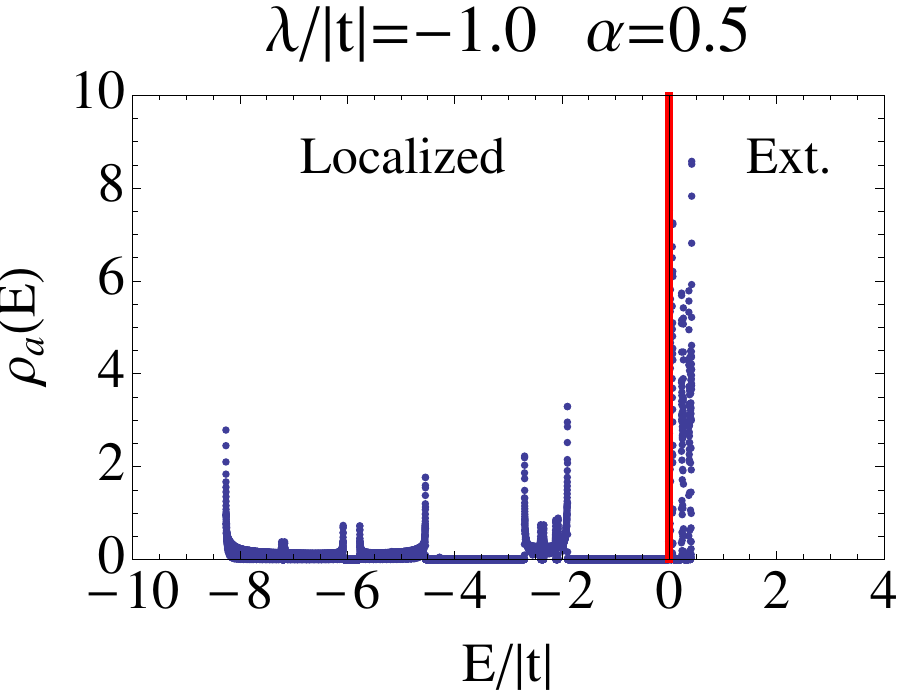}
  \end{minipage}
  \begin{minipage}{0.3\textwidth}
\centering
  \includegraphics[width=4.5cm]{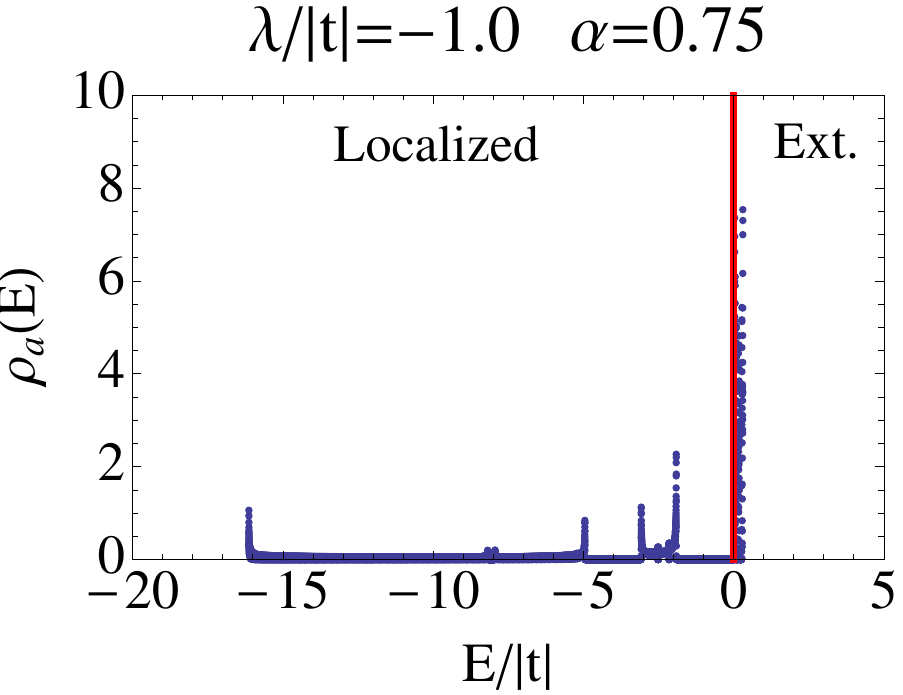}
  \end{minipage}
    \begin{minipage}{0.3\textwidth}
\centering
  \includegraphics[width=4.5cm]{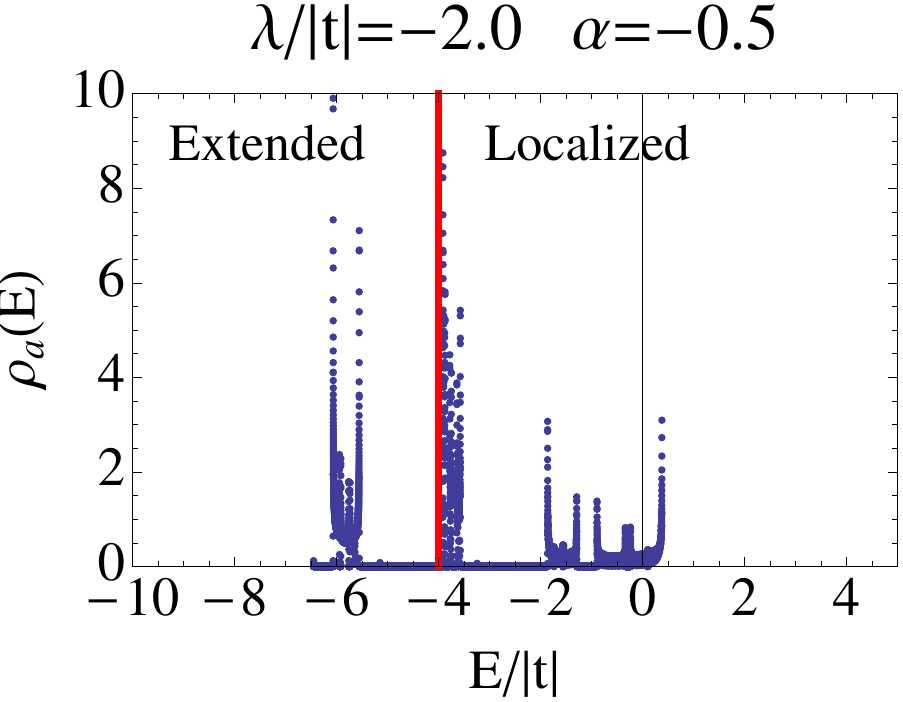}
  \end{minipage}
  \begin{minipage}{0.3\textwidth}
\centering
  \includegraphics[width=4.5cm]{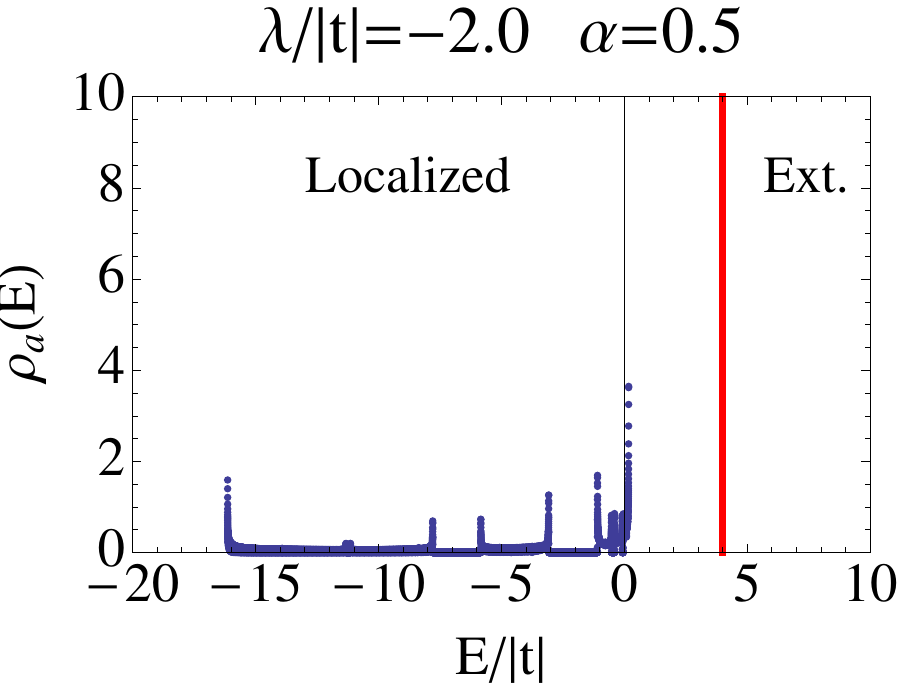}
  \end{minipage}
  \begin{minipage}{0.3\textwidth}
\centering
  \includegraphics[width=4.5cm]{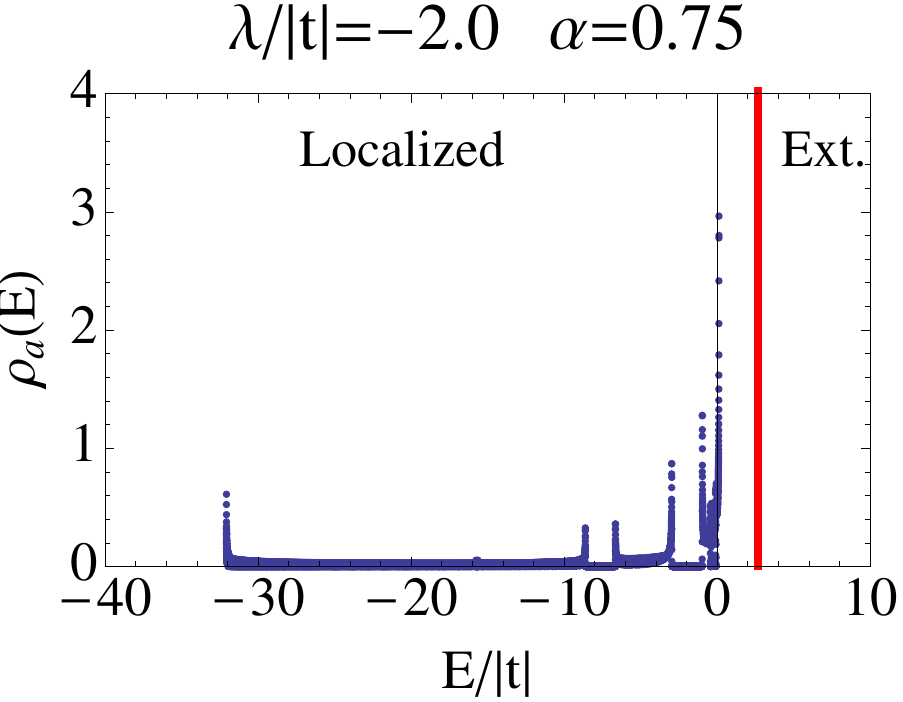}
  \end{minipage}
   \begin{minipage}{0.3\textwidth}
\centering
  \includegraphics[width=4.5cm]{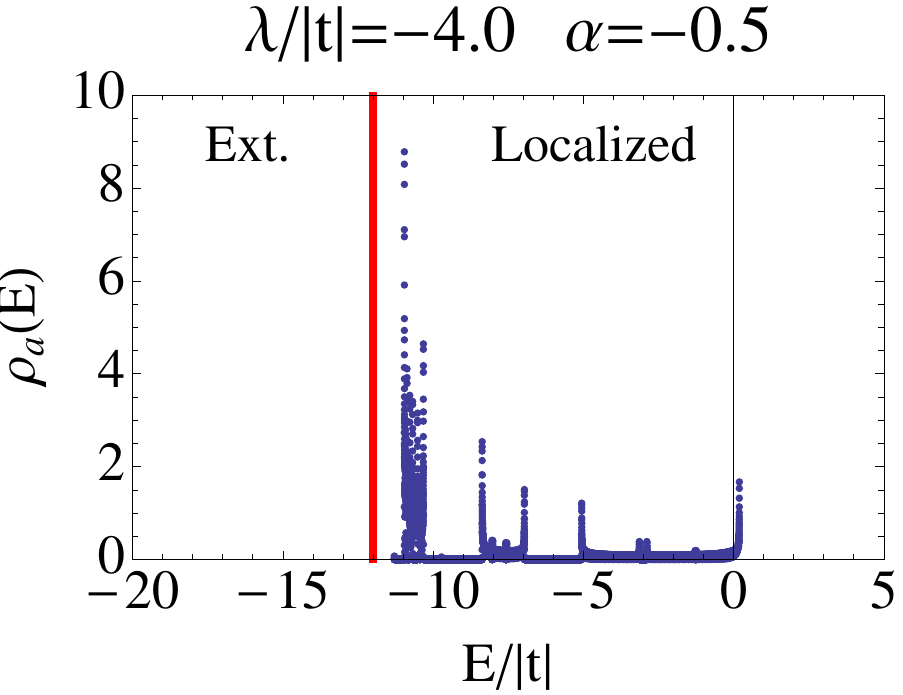}
  \end{minipage}
  \begin{minipage}{0.3\textwidth}
\centering
  \includegraphics[width=4.5cm]{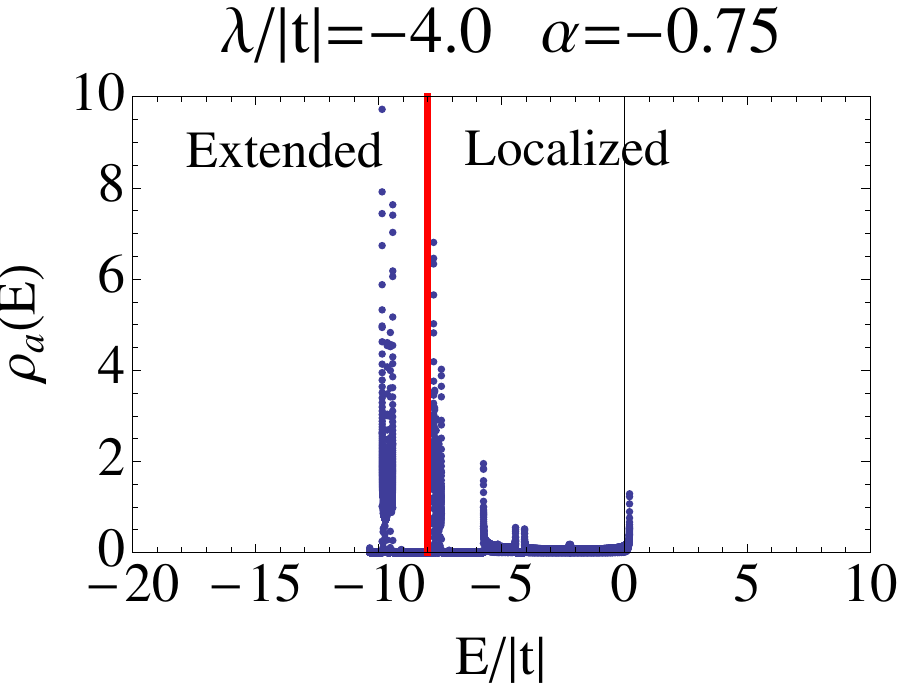}
  \end{minipage}
  \begin{minipage}{0.3\textwidth}
\centering
  \includegraphics[width=4.5cm]{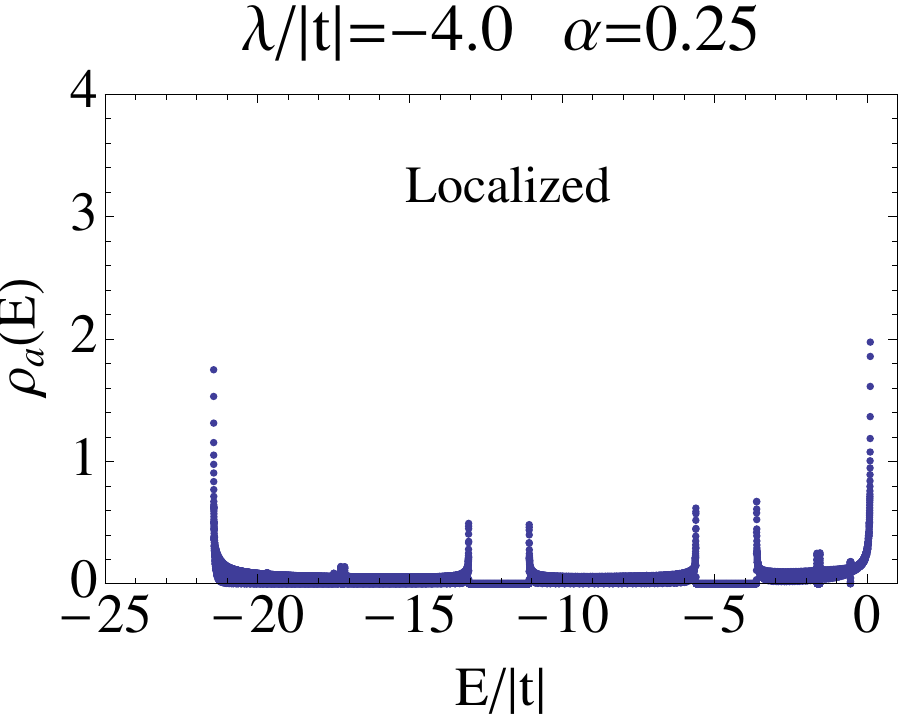}
  \end{minipage}
\caption{Density of states as a function of energy $E$ for the potential defined in Eq. (\ref{onsite2}) for various values of $\lambda$ and $\alpha$ as specified within each figure.  We also show the analytic result for the location of the mobility edge in red.}
\label{fig:rhot_a2}
\end{figure}

 As the KPM directly evaluates the local and average density of states, it does not determine the energy eigenvalues.  Therefore, in order to plot the energy and $\alpha$ dependence while distinguishing between gapped and localized states, we filter the typical density of states based on the average density of states.  In some regions in the energy and $\alpha$ phase diagram the density of states can become very small, resulting in us not plotting any value for the typical density of states, even though there is a finite eigenvalue (but with a really small weight) in that region.  To show this clearly, and also display where the localization transition occurs relative to the regions of the overall bandwidth, we present various cuts of the average density of states in Figs.~\ref{fig:rhot_a1} and \ref{fig:rhot_a2}, while showing the location of the analytic result for the mobility edge in red.
  \section{Typical Density of States compared with Inverse participation Ratio}
  %%%%%%%%%%%%%%%%%%%%%%%%%%%%%%%%%%%%%%%%%%%%%%%%%%%%%%%%%%%%%%%%%%%%%%%%%%%%%%%%%%%%%%%%
 \begin{figure}[htb!]
  \centering
\includegraphics[width=4cm,height=3.5cm]{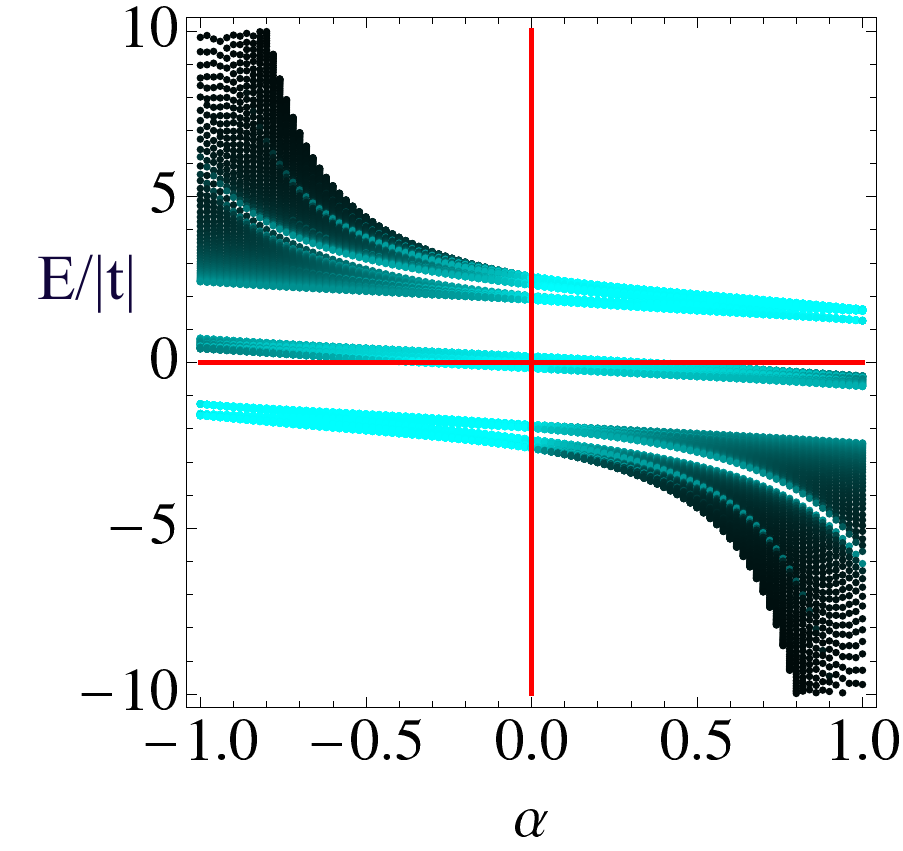}\hspace{0.1cm}\includegraphics[width=4.0cm,height=3.5cm]{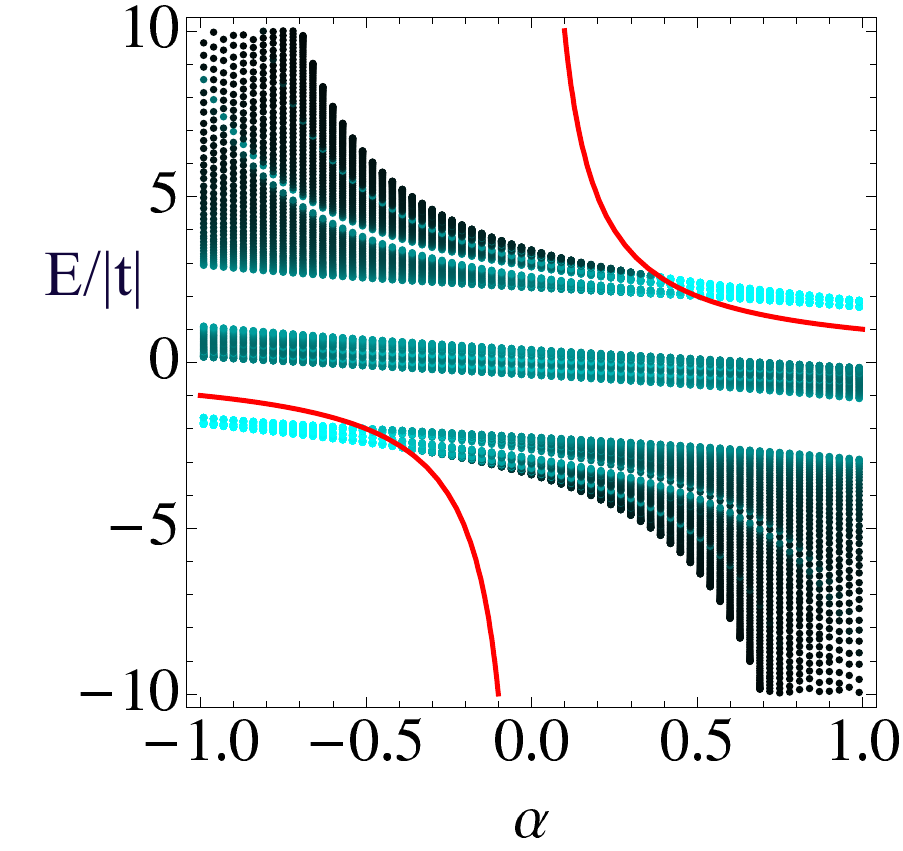}\\
\hspace{0.7cm}\textrm{(a) IPR}\hspace{4cm}\textrm{(b) IPR}\\
\includegraphics[width=4.0cm,height=3.5cm]{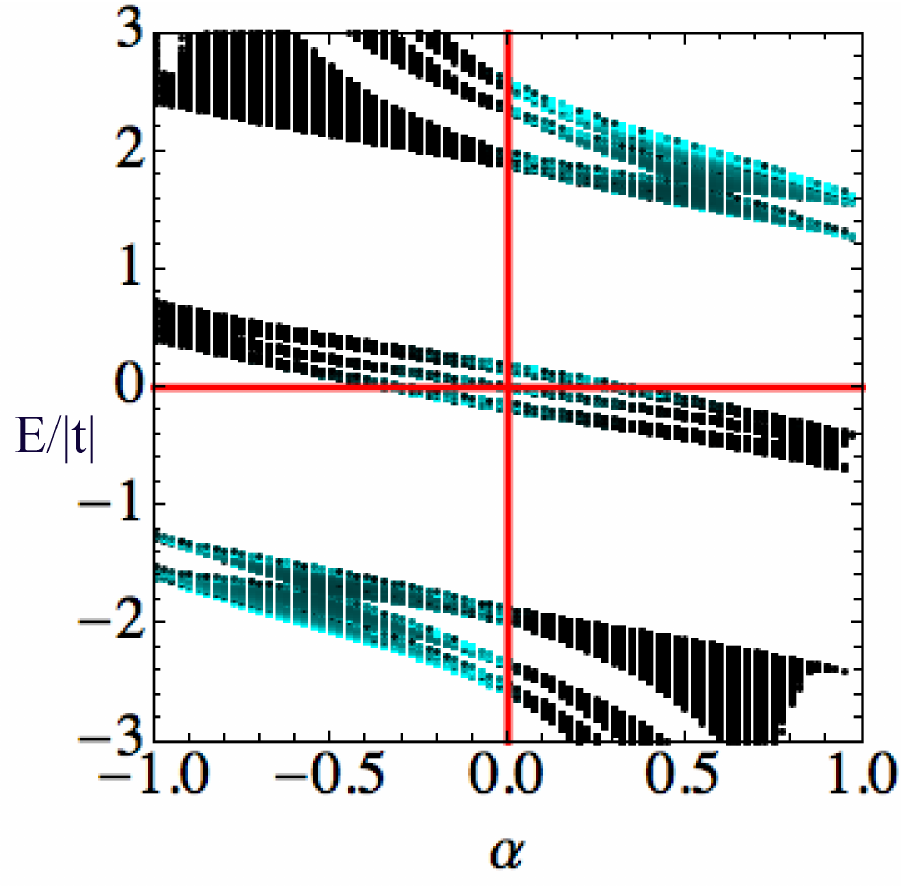}\hspace{0.1cm}\includegraphics[width=4.0cm,height=3.5cm]{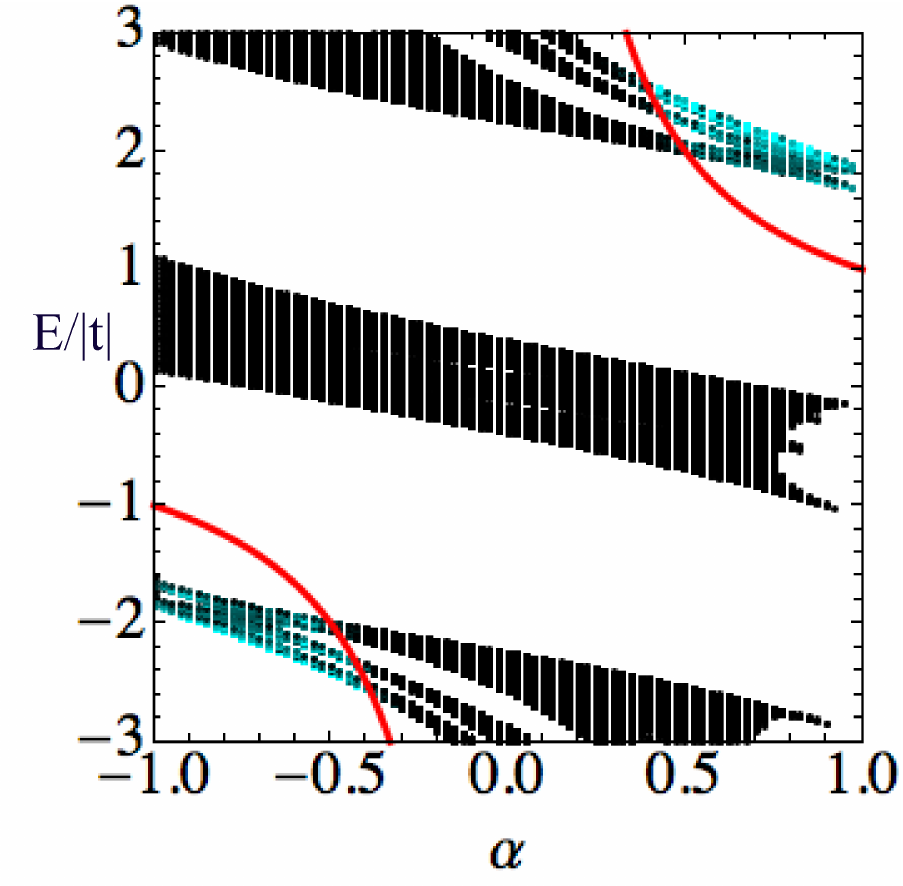}\\
\hspace{0.7cm}\textrm{(c) TDOS}\hspace{4cm}\textrm{(d) TDOS}\\
%\includegraphics[width=4.0cm,height=3.5cm]{fig2f.pdf}\hspace{0.1cm}\includegraphics[width=4.0cm,height=3.5cm]{fig2fS.pdf}\\
%\hspace{0.7cm}\textrm{(c)}\hspace{4cm}\textrm{(d)}\\
\vspace{0.1cm}
\includegraphics[width=6.0cm,height=0.8cm]{figlegend.pdf}
    \caption{Results for the onsite potential in Eq. (\ref{onsite}). (a, b): Numerical energy spectrum $E/|t|$ with $t=-1.0$ as a function of $\alpha$ for $L=500$ sites for  $\lambda/|t|=-1.0$ and $\lambda/|t|=-1.5$. Pure cyan denotes IPR=0 and pure black denotes IPR=1.  (c, d): TDOS plotted after filtering based on the DOS for $L=10,000$ and an expansion order $N_c=32,768$ for  $\lambda/|t|=-1.0$ and $\lambda/|t|=-1.5$. Pure cyan denotes maximum TDOS values between 1 and 10 for an extended state and pure black denotes TDOS=0 for a localized state.  Eigenvalues that have a very small DOS are not plotted, hence the discrepancy in the black regions between the TDOS and the IPR. Red line is a plot of the analytically obtained critical condition defined in Eq.~(\ref{mainresult}).}  
     \la{ipr_dual1}
 \end{figure}
%%%%%%%%%%%%%%%%%%%%%%%%%%%%%%%%%%%%%%%%%%%%%%%%%%%%%%%%%%%%%%%%%%%%% %%
\begin{figure}
\centering
\begin{minipage}{0.4\textwidth}
\centering
  \includegraphics[width=6cm]{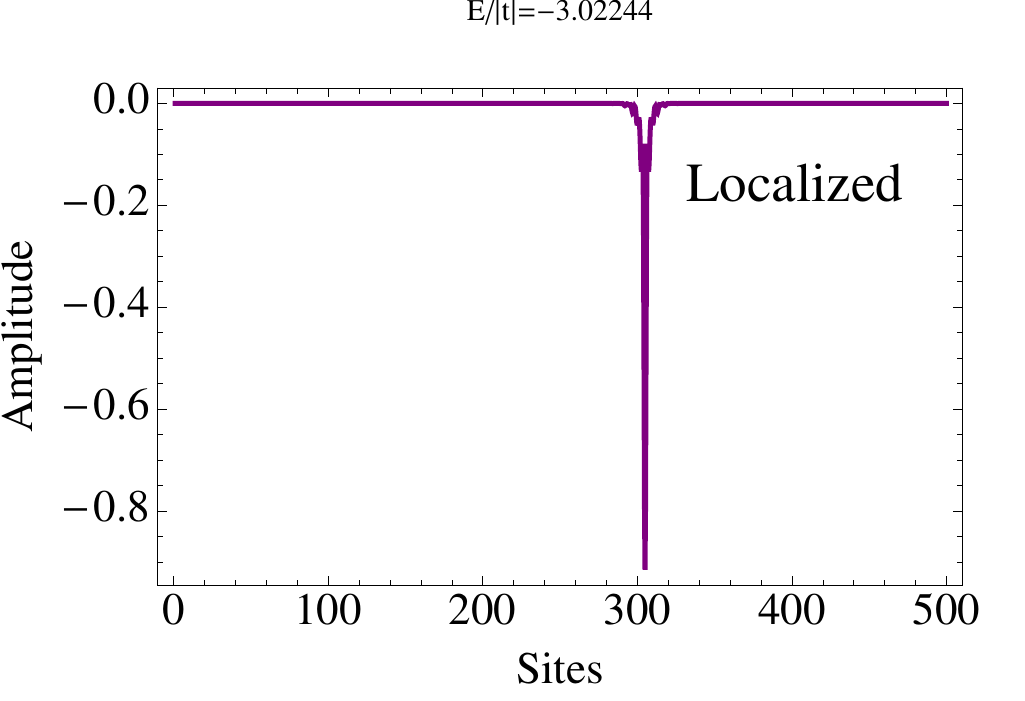}
  \end{minipage}
  \begin{minipage}{0.4\textwidth}
\centering
  \includegraphics[width=6cm]{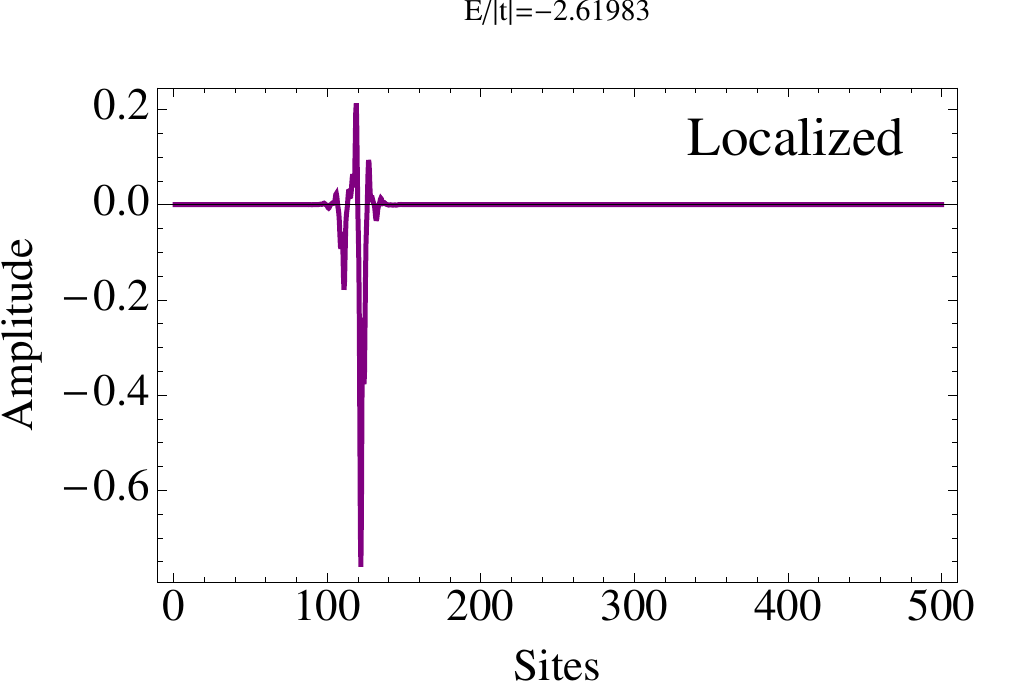}
  \end{minipage}
  \begin{minipage}{0.4\textwidth}
\centering
  \includegraphics[width=6cm]{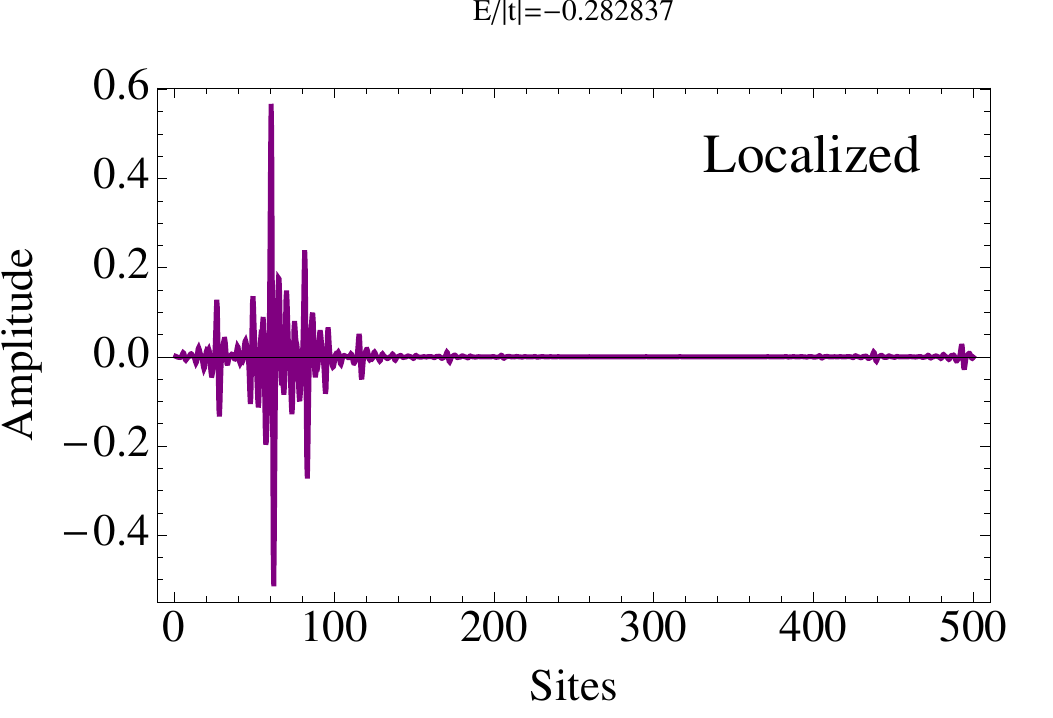}
  \end{minipage}
  \begin{minipage}{0.4\textwidth}
\centering
  \includegraphics[width=6cm]{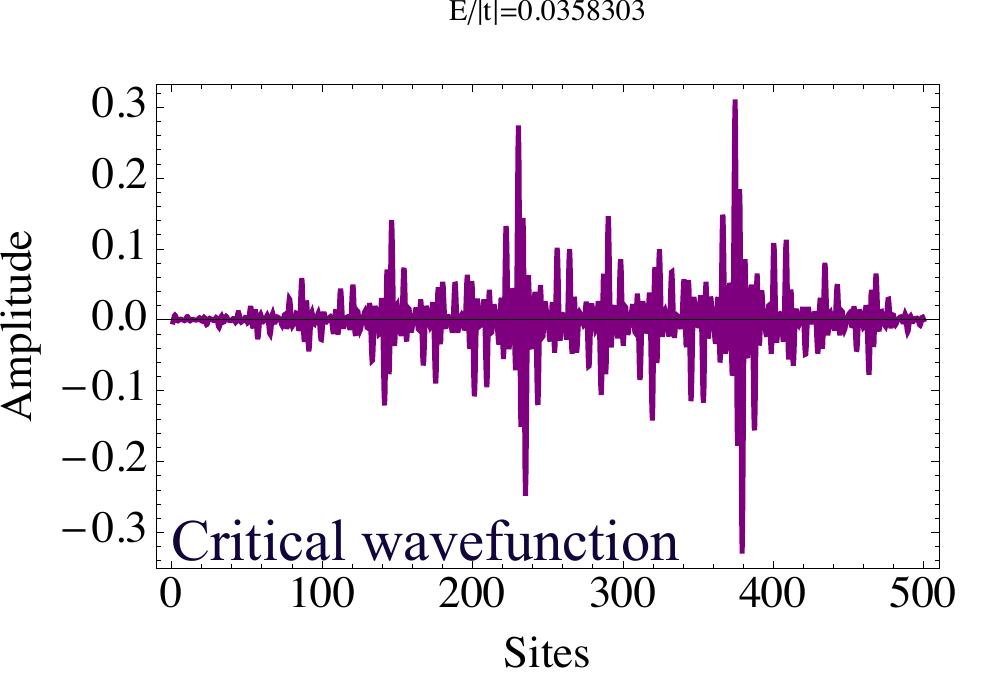}
  \end{minipage}
  \begin{minipage}{0.4\textwidth}
\centering
  \includegraphics[width=6cm]{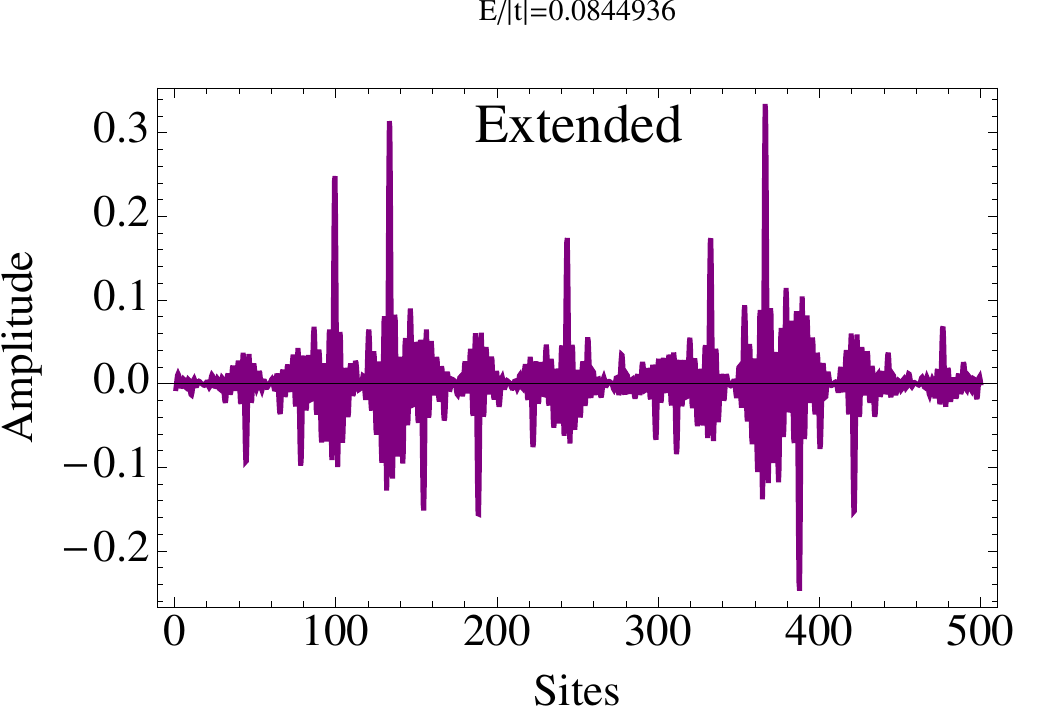}
  \end{minipage}
  \begin{minipage}{0.4\textwidth}
\centering
  \includegraphics[width=6cm]{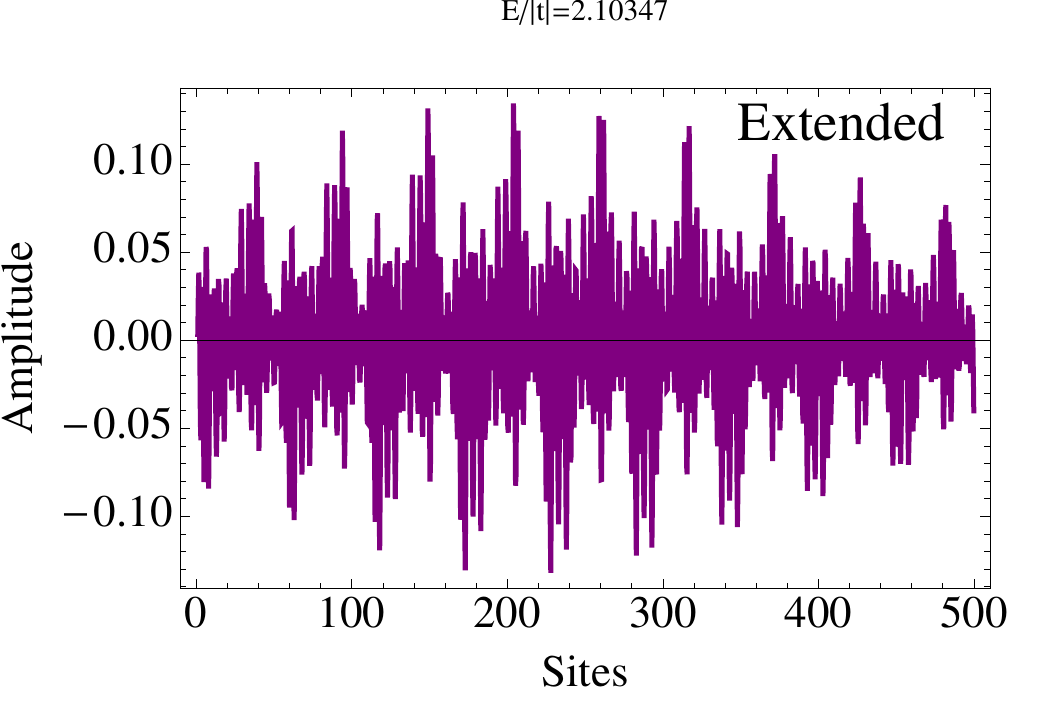}
  \end{minipage}
\caption{Wave function plot for different energies $E/|t|$ (shown in the figure) for the potential defined in Eq. (\ref{onsite}) for $\lambda/|t|=-1.0$ and $\alpha=0.2$. The IPR and TDOS plots for the corresponding parameters are given in Figs.~\ref{ipr_dual1}~a and  c respectively. }
\label{fig:wf}
\end{figure}
%%%%%%%%%%%%%%%%%%%%%%%%%%%%%%%%%%%%%%%%%%%%%%%%%%%%%%%%%%%%%%%%%%%%%%%%%%%%%%%%%%%%%%
 \begin{figure}[htb!]
  \centering
\includegraphics[width=4cm,height=3.5cm]{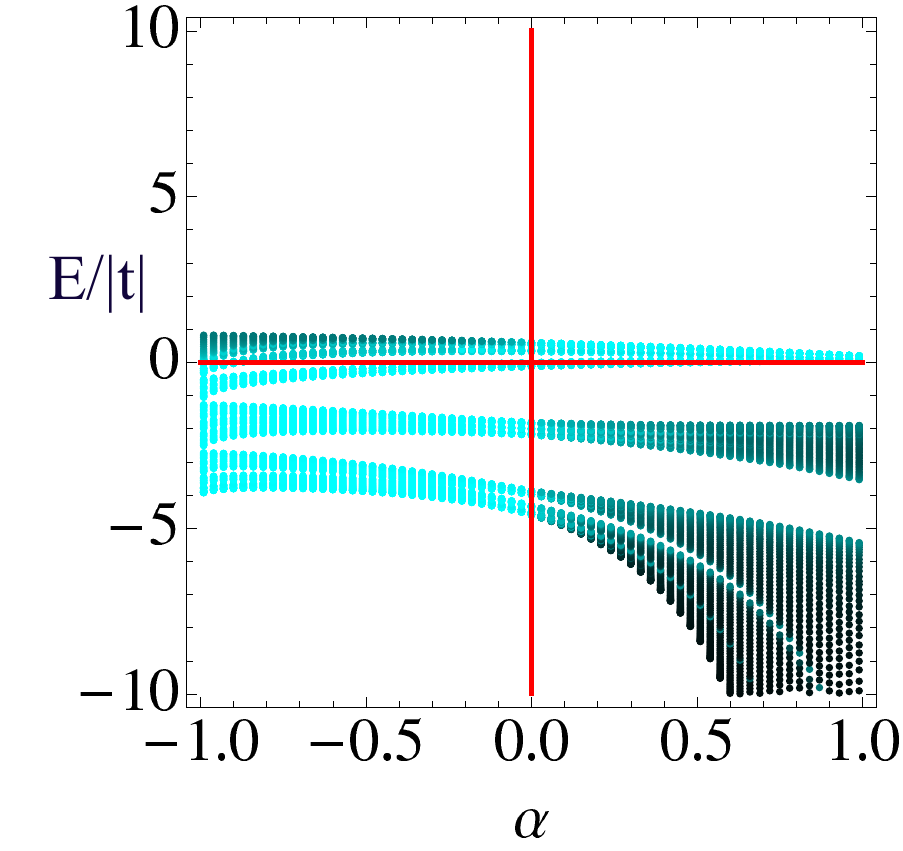}\hspace{0.1cm}\includegraphics[width=4.0cm,height=3.5cm]{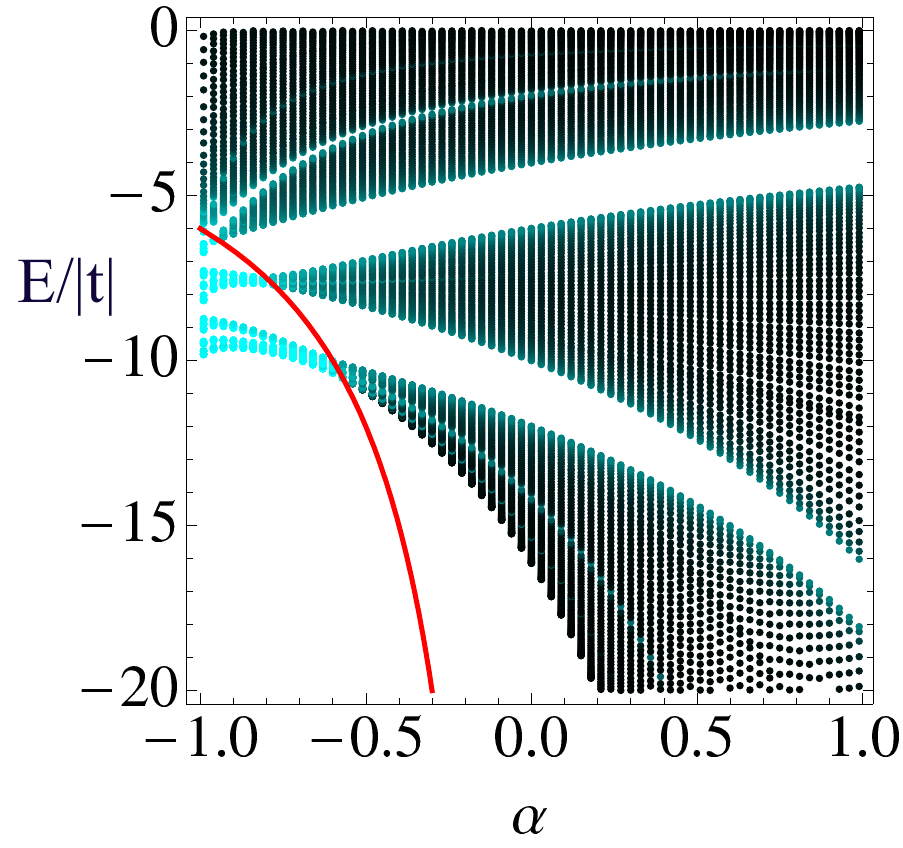}\\
\hspace{0.7cm}\textrm{(a) IPR}\hspace{4cm}\textrm{(b) IPR}\\
\includegraphics[width=4.0cm,height=3.5cm]{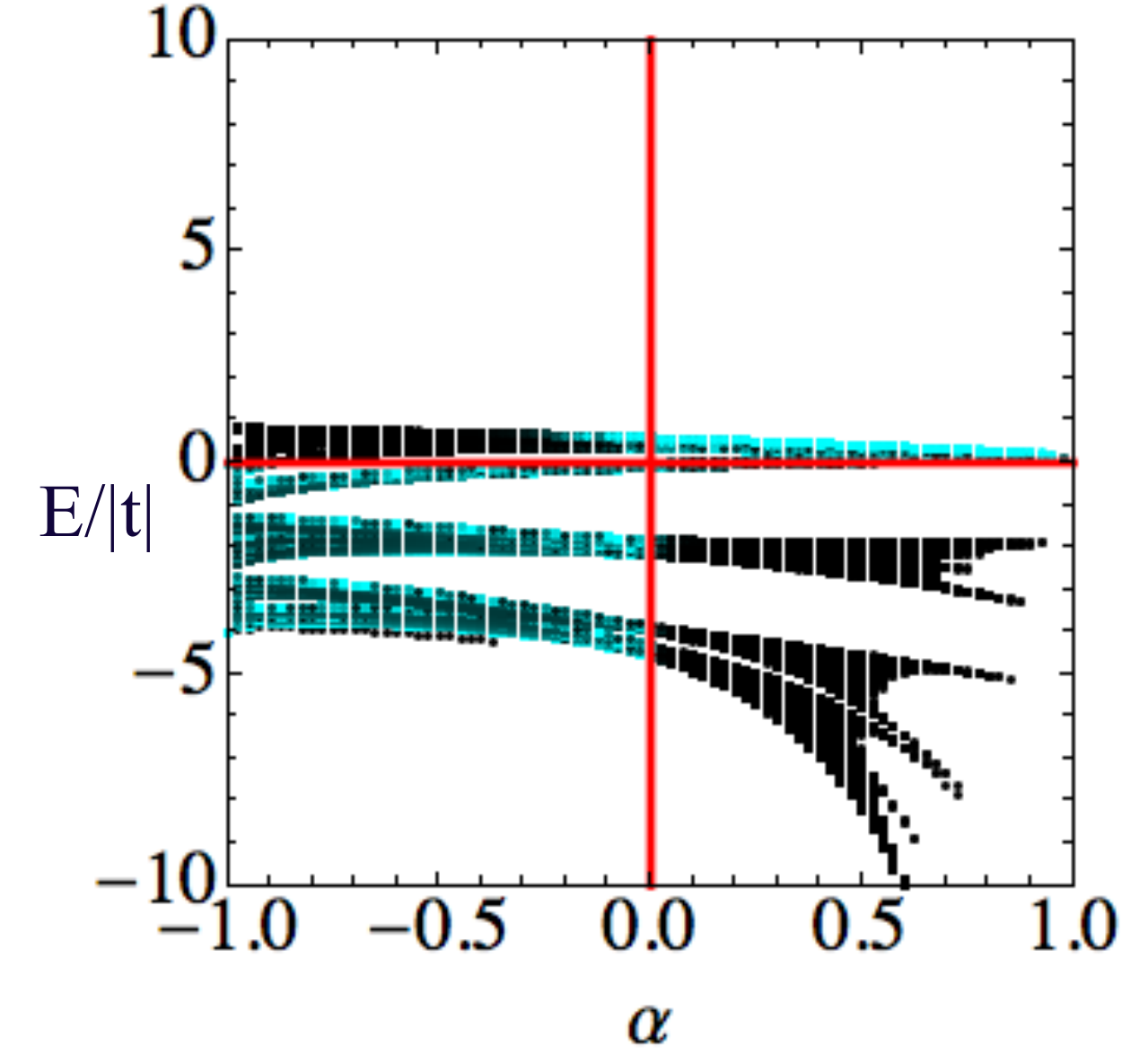}\hspace{0.1cm}\includegraphics[width=4.0cm,height=3.5cm]{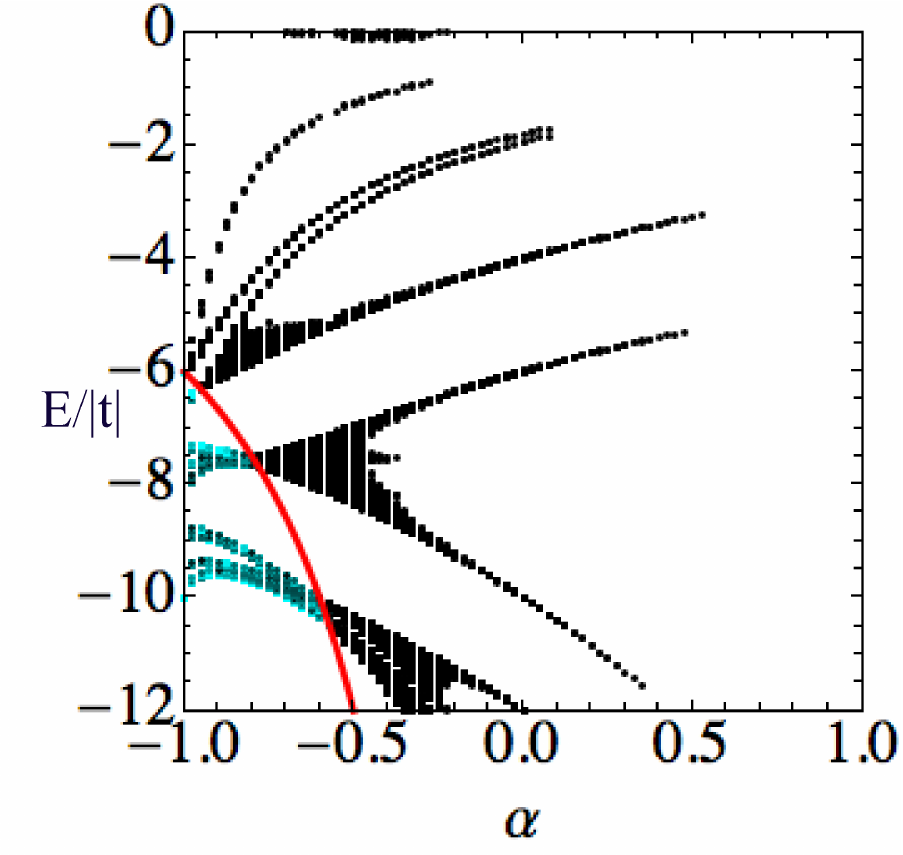}\\
\hspace{0.7cm}\textrm{(c) TDOS}\hspace{4cm}\textrm{(d) TDOS}\\
\vspace{0.1cm}
\includegraphics[width=6.0cm,height=0.8cm]{figlegend.pdf}
    \caption{Results for the onsite potential in Eq. (\ref{onsite2}). (a, c): Numerical energy spectrum $E/|t|$ with $t=-1.0$ as a function of $\alpha$ for $L=500$ sites for  $\lambda/|t|=-1.0$ and $\lambda/|t|=-4.0$. Pure cyan denotes IPR=0 and pure black denotes IPR=1.  (b, d): TDOS plotted after filtering based on the DOS for $L=10,000$ and an expansion order $N_c=32,768$ for $\lambda/|t|=-1.0$ and $\lambda/|t|=-4.0$. Pure cyan denotes maximum TDOS values between 1 and 10 for an extended state and pure black denotes TDOS=0 for a localized state. Eigenvalues that have a very small DOS are not plotted, hence the discrepancy in the black regions between the TDOS and the IPR. Red line is a plot of the analytically obtained critical condition defined in Eq.~(\ref{mainresult}).}  
  \la{ipr_dual2}
\end{figure}
 %%%%%%%%%%%%%%%%%%%%%%%%%%%%%%%%%%%%%%%%%%%%%%%%%%%%%%%%%%%%%%%%%%%%  
 \begin{figure}[htb!]
\centering
\begin{minipage}{0.4\textwidth}
\centering
  \includegraphics[width=6cm]{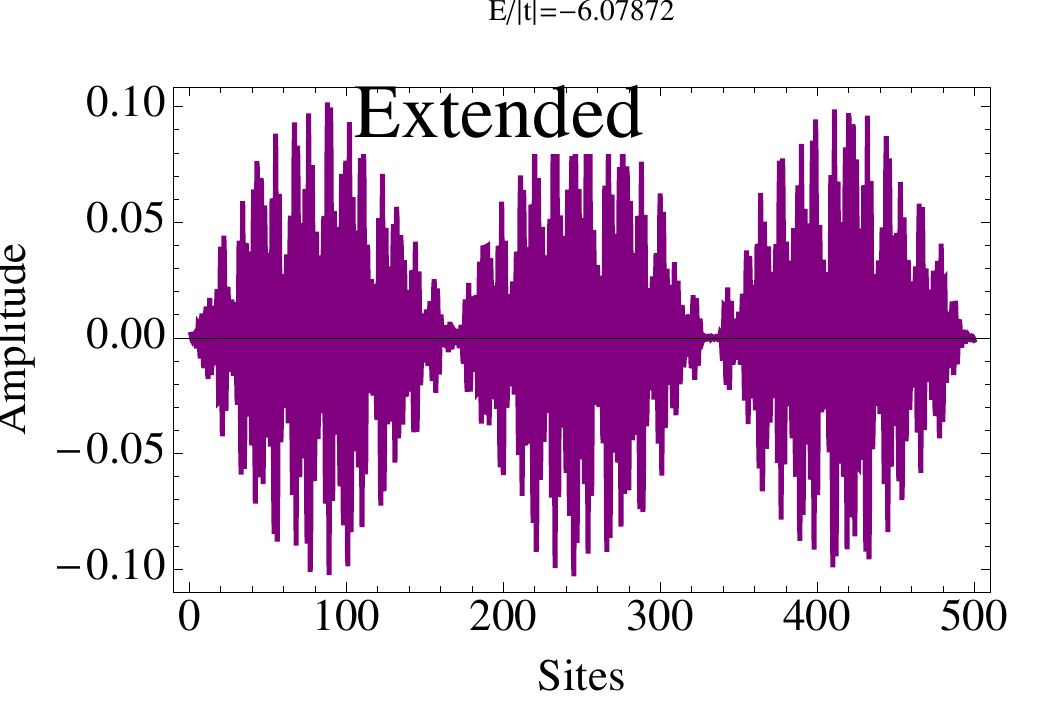}
  \end{minipage}
  \begin{minipage}{0.4\textwidth}
\centering
  \includegraphics[width=6cm]{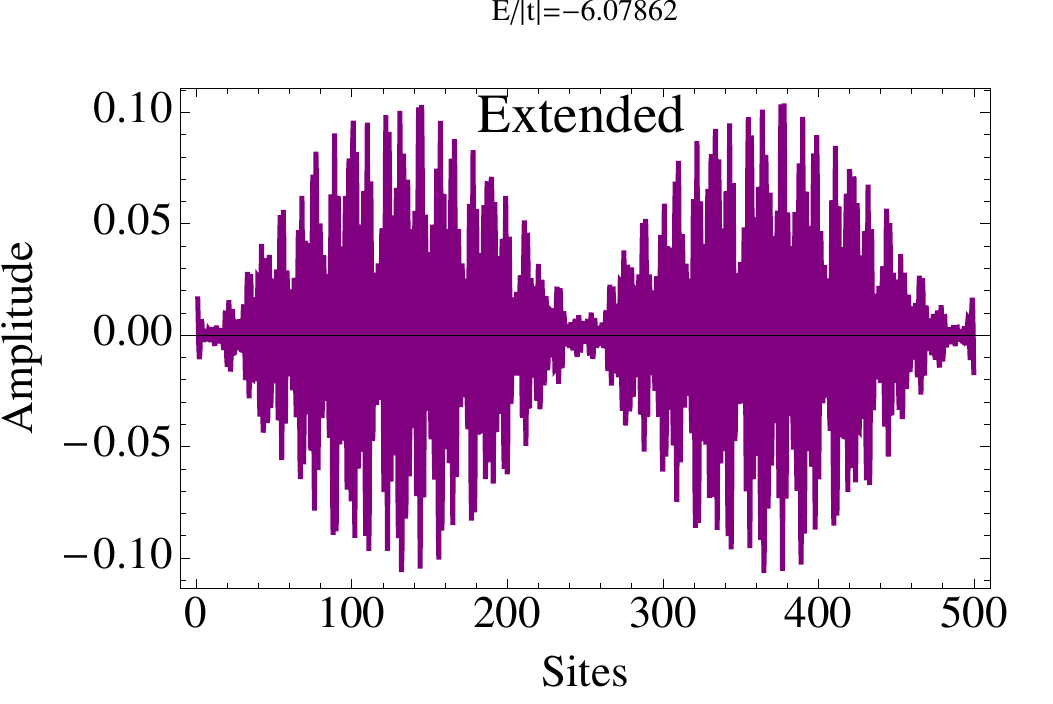}
  \end{minipage}
  \begin{minipage}{0.4\textwidth}
\centering
  \includegraphics[width=6cm]{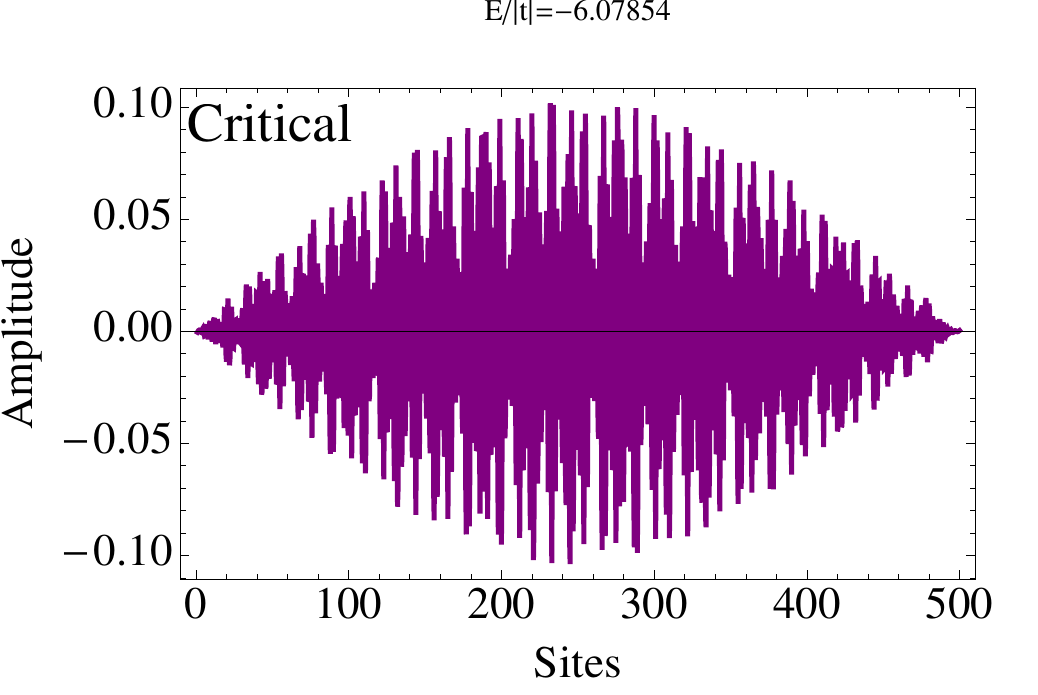}
  \end{minipage}
  \begin{minipage}{0.4\textwidth}
\centering
  \includegraphics[width=6cm]{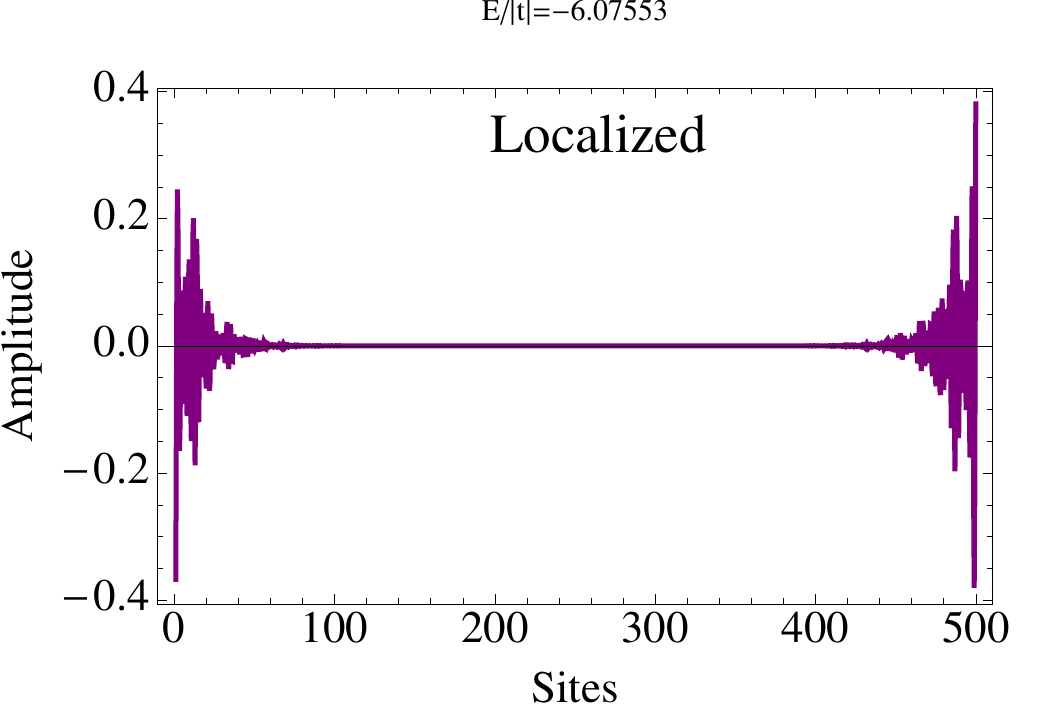}
  \end{minipage}
  \begin{minipage}{0.4\textwidth}
\centering
  \includegraphics[width=6cm]{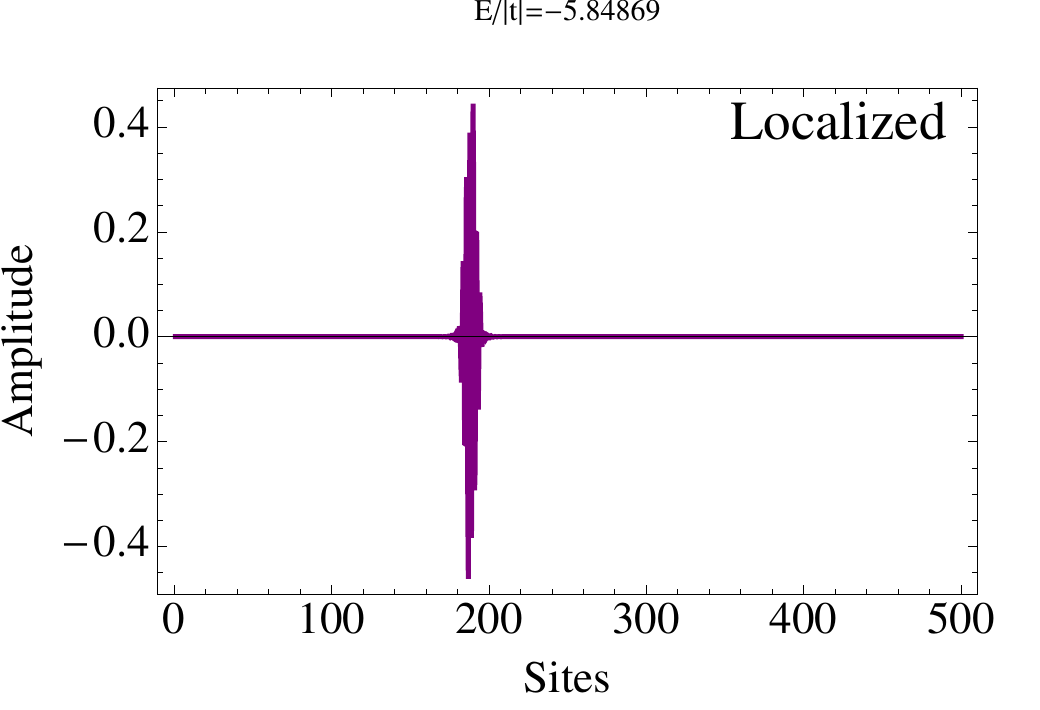}
  \end{minipage}
  \begin{minipage}{0.4\textwidth}
\centering
  \includegraphics[width=6cm]{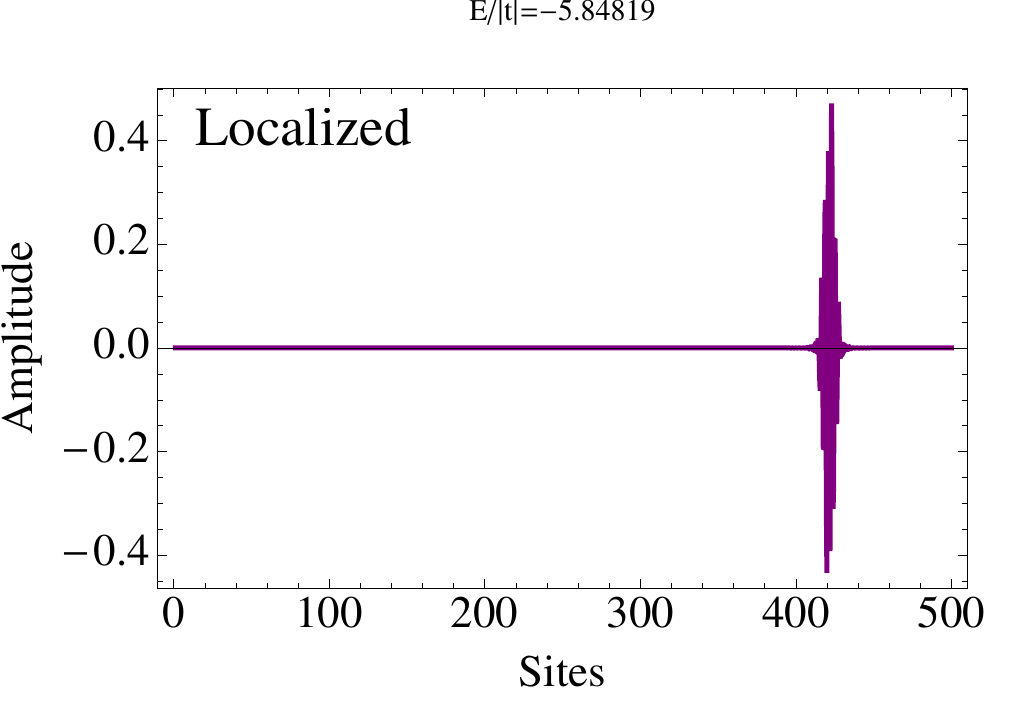}
  \end{minipage}
\caption{Wave function plot for different energies $E/|t|$ (shown in the figure) for the potential defined in Eq. (\ref{onsite2}) for $\lambda/|t|=-4.0$ and $\alpha=-0.99$. The IPR and TDOS plots for the corresponding parameters are given in Figs.~\ref{ipr_dual2}~b and  d respectively. }
\label{fig:wf2}
\end{figure}
%%%%%%%%%%%%%%%%%%%%%%%
  In this section, we compare the localization properties of the two models discussed in the main text for different parameter values. The first onsite potential for the nearest neighbor tight binding model considered in the main text is defined as, 
\begin{align}
V_n(\alpha,\phi)=2\lambda\frac{\cos (2\pi n b+\phi)}{1-\alpha\cos (2\pi n b+\phi)}.\la{onsite}
\end{align} 
For a quasiperiodic modulation, we set $b$ to be irrational (we choose $1/b=\frac{\sqrt{5}-1}{2}$ for our numerical work). The mobility edge separating the localized and extended states for Eq.~(\ref{onsite}) is given by the following expression,
\begin{align}
\alpha E= 2 \sgn(\lambda)(|t|-|\lambda|). \la{mainresult}
\end{align}
We superimpose the IPR and TDOS result with the critical condition conjectured in Eq.~(\ref{mainresult}).  Fig.~(\ref{ipr_dual1}a, c) is a plot of energy eigenvalue $E/|t|$ with $t=-1.0$ as a function of $\alpha$ for fixed values of $\lambda/|t|$.  In Fig.~(\ref{ipr_dual1}a and b), we fix $\lambda/|t|=-1.0$, which is the critical condition for the AAH model. We show that the critical line is formed by $\alpha=0$ and $E=0$ which divides the $(\alpha, E/|t|)$ space into four quadrants. The IPR indicates a localized (extended) to extended (localized) transition across the critical lines in excellent agreement with the analytical critical boundary defined by Eq.~(\ref{mainresult}). In Figs.~\ref{ipr_dual1}c, d we fix $\lambda/|t|=-1.5$. The $\alpha=0$ slice of these plots correspond to the AAH model with all the states being localized. For $\alpha \ne 0$, the states remain localized till they encounter the mobility edge in the $(\alpha, E/|t|)$ space. Across this mobility edge the states undergo a localization transition to extended states in agreement with the IPR and the TDOS results. We verify from two independent numerical methods (i. e. IPR and TDOS) that the analytical mobility edge is indeed the critical condition for the localization transition.
To understand the nature of the wave function across the mobility edge we plot the wavefunction as a function of position for different values of energy eigenvalues for $\alpha=0.2$ and $\lambda/|t|=-1.0$. The critical energy for this case is given by $E/|t|=0$. According to the analytical condition, all the negative energy eigenstates are localized and all the positive energy eigenstates are extended. The energy states close to zero energy are critical in nature. Wavefunction plots in Figs.~(\ref{fig:wf}) is in excellent agreement with the analytical prediction from the duality condition, IPR and TDOS.

  The second onsite potential for the nearest neighbor tight binding model considered in the main text is defined as, 
\bea
V_{n}(\alpha,\phi)&=&2\lambda\frac{1-\cos(2\pi n b+\phi)}{1+\alpha\cos(2\pi n b+\phi)}.
\la{onsite2}
\eea

As shown in the main text, there are a whole family of models that satisfy the mobility edge condition defined in Eq.~\ref{mainresult}.  In Fig.~(\ref{ipr_dual2}), we plot the numerical spectrum as a function of $\alpha$ with a color coded IPR (Fig.~\ref{ipr_dual2}a, c) and TDOS (Fig.~\ref{ipr_dual2}b, d) for $\lambda/|t|=-1.0, \  -4.0$. The critical condition (shown in red line ) is in excellent agreement with the separation of the localized and extended states as indicated by the IPR and TDOS. Figs.~(\ref{fig:wf2}) show plots of the wavefunction as a function of position for different values of energy eigenvalues in the neighborhood of the critical energy for $\alpha=-0.99$ and $\lambda/|t|=-4.0$. The critical energy for this case is given by $E/|t|=-6.07$. According to the analytical condition, all the energy eigenstates below the critical energy are extended and all the energy eigenstates are above the critical energy extended. Figs.~(\ref{fig:wf})  is in excellent agreement with the analytical prediction from the duality condition, IPR and TDOS.

 \section{ Density of States}
 In this section we present the average density of states in Eq. (\ref{eqn:rhoa}) for both models defined by the onsite potentials Eqs. (\ref{onsite}) and (\ref{onsite2}), calculated using the KPM.  For the calculations presented here we consider $L=10,000$ sites and keep $N_c=32,768$ moments.

\end{document}